\documentclass{aa} 

\usepackage[varg]{txfonts}
\usepackage{graphicx}
\usepackage{longtable}
\usepackage{lscape}
\usepackage{multirow}
\usepackage{microtype}
\usepackage[none]{hyphenat}

\newcommand{\kms}{\mbox{km~s$^{-1}$}}

\newcommand{\fei}{Fe\,{\scriptsize I}}
\newcommand{\sii}{Si\,{\scriptsize I}}
\newcommand{\vi}{V\,{\scriptsize I}}
\newcommand{\feii}{Fe\,{\scriptsize II}}

\def\feh{\hbox{[\rm{Fe/H}]}}
\def\feih{\hbox{[\rm{Fe\,{\scriptsize I}/H}]}}
\def\feiih{\hbox{[\rm{Fe\,{\scriptsize II}/H}]}}

\def\Rg{R\rm_G}
\def\logg{log $g$}
\def\vt{$v_t$}
\def\teff{$T\rm_{eff }$}
\def\snr{$S/N$}

\begin{document}

\title{On the fine structure of the Cepheid metallicity gradient in the 
Galactic thin disk\thanks{Based on spectra collected with the spectrograph 
UVES available at the ESO Very Large Telescope (VLT), Cerro Paranal, 
(081.D-0928(A) PI: S.~Pedicelli -- 082.D-0901(A) PI: S.~Pedicelli 
-- 089.D-0767 PI: K. Genovali).}}

\author{K. Genovali\inst{1}
\and B. Lemasle\inst{2} 
\and G. Bono\inst{1,3} 
\and M. Romaniello\inst{4} 
\and M. Fabrizio\inst{5} 
\and I. Ferraro\inst{3} 
\and G. Iannicola\inst{3} 
\and C.D. Laney\inst{6,7}
\and M. Nonino\inst{8} 
\and M. Bergemann\inst{9,10}  
\and R. Buonanno\inst{1,5}
\and P. Fran\c cois\inst{11,12}
\and L. Inno\inst{1,4}
\and R.-P. Kudritzki\inst{13,14,9}  
\and N. Matsunaga\inst{15}
\and S. Pedicelli\inst{1}
\and F. Primas\inst{4}
\and F. Th\'evenin\inst{16}
}
\institute{
Dipartimento di Fisica, Universit\`{a} di Roma Tor Vergata, via della Ricerca Scientifica 1, 
00133 Rome, Italy \email{katia.genovali@roma2.infn.it}
\and Astronomical Institute `Anton Pannekoek', Science Park 904, P.O. Box 94249, 1090 GE Amsterdam, The Netherlands
\and INAF--Osservatorio Astronomico di Roma, via Frascati 33, Monte Porzio Catone, Rome, Italy
\and European Southern Observatory, Karl-Schwarzschild-Str. 2, D-85748 Garching bei Munchen, Germany
\and INAF-Osservatorio Astronomico di Collurania, via M. Maggini, I-64100 Teramo, Italy
\and Department of Physics and Astronomy, N283 ESC, Brigham Young University, Provo, UT 84601, USA
\and South African Astronomical Observatory, P.O. Box 9, Observatory 7935, South Africa
\and INAF--Osservatorio Astronomico di Trieste, via G.B. Tiepolo 11, 40131 Trieste, Italy 
\and Max-Planck-Institut fur Astrophysik, Karl-Schwarzschild-Str. 1, 85741 Garching, Germany 
\and Institute of Astronomy, University of Cambridge, Madingley Road, CB3 0HA, Cambridge, United Kingdom
\and GEPI - Observatoire de Paris, 64 Avenue de l'Observatoire, 75014 Paris, France
\and UPJV - Universit\'e de Picardie Jules Verne, 80000 Amiens, France
\and Institute for Astronomy, University of Hawai’i, 2680 Woodlawn Dr, Honolulu, HI 96822, USA
\and University Observatory Munich, Scheinerstr. 1, D-81679 Munich, Germany
\and Kiso Observatory, Institute of Astronomy, School of Science, The University of Tokyo 10762-30, Mitake, Kiso-machi, Kiso-gun, 3 Nagano 97-0101, Japan
\and Laboratoire Lagrange, UMR7293, Universit\'e de Nice Sophia-Antipolis, CNRS, Observatoire de la C\^ote d'Azur, 06300 Nice, France
}
\date{Received <date> / Accepted <date>}

\abstract{
We present homogeneous and accurate iron abundances for 42 Galactic Cepheids 
based on high–spectral resolution (R$\sim$ 38,000) high signal-to-noise ratio 
(SNR $\geq$ 100) optical spectra collected with UVES at VLT (128 spectra). 
The above abundances were complemented with high--quality iron abundances 
provided either by our group (86) or available in the literature. We paid 
attention in deriving a common metallicity scale and ended up with 
a sample of 450 Cepheids. We also estimated for the entire sample accurate 
individual distances by using homogeneous near-infrared photometry 
and the reddening free Period-Wesenheit relations.  
The new metallicity gradient is linear over a broad range of Galactocentric 
distances ($\Rg \sim$5--19 kpc) and agrees quite well with similar estimates 
available in the literature (-0.060$\pm$0.002 dex/kpc). 
We also uncover evidence which suggests that the residuals of the metallicity 
gradient are tightly correlated with candidate Cepheid Groups (CGs). The 
candidate CGs have been identified as spatial overdensities of Cepheids 
located across the thin disk. They account for a significant fraction 
of the residual fluctuations, and in turn for the large intrinsic 
dispersion of the metallicity gradient.   
We performed a detailed comparison with metallicity gradients based 
on different tracers: OB stars and open clusters. 
We found very similar metallicity gradients for ages younger than 
3 Gyrs, while for older ages we found a shallower slope and an increase 
in the intrinsic spread. The above findings rely on homogeneous age, 
metallicity and distance scales.  
Finally we found, by using a large sample of Galactic and Magellanic 
Cepheids for which are available accurate iron abundances, that the 
dependence of the luminosity amplitude on metallicity is vanishing.  
}


\titlerunning{The fine structure of the metallicity gradient}
\authorrunning{Genovali et al.}

\maketitle

\section{Introduction}

Recent findings concerning the metallicity gradient across the Galactic thin disk, 
based on high spectral resolution, high signal-to-noise spectra and on stellar 
tracers for which we can provide accurate individual Galactocentric distances, 
are disclosing a new interesting scenario. 
The iron gradients traced by stellar populations younger than a few hundred 
of Myrs show a well defined trend when moving from the inner to the outer 
disc regions. The iron abundances in the innermost disc regions ($R_G\sim$5 kpc) 
are well above solar 
\citep[\feh $\sim$0.4,][hereinafter G13]{Andrievsky2002,Pedicelli2009,Luck2011a,Luck2011b,Genovali2013} 
while in the outer disk ($R_G\sim$15 kpc) 
they are significantly more metal--poor 
\citep[\feh $\sim$-0.2/-0.5,][]{Andrievsky2004,Yong2006,Lemasle2008}. 
However, the young stellar populations in the two innermost Galactic regions showing 
ongoing star formation activity --the Bar and the Nuclear Bulge-- attain solar
iron abundances. Thus suggesting that the above regions are experiencing  
different chemical enrichment histories \citep{Bono2013}.

The use of high-quality data and homogeneous analysis of large sample of classical 
Cepheids and young massive main sequence stars provided the opportunity to 
overcome some of the systematics affecting early estimates of the metallicity 
gradient. However, current findings still rely on several assumptions that 
might introduce systematic errors. 

i) Distances -- Cepheids are very solid primary distance indicators, but
they only trace young stellar populations. The use of red clump stars
is very promising, since they are ubiquitous in the innermost Galactic
regions. However, their individual distances might be affected by larger
uncertainties, since we are dealing with stellar populations covering
a broad range in ages and in metal abundances \citep{Girardi2001,Salaris2002}.

ii) Gradients -- Recent spectroscopic investigations indicate that the
use of homogeneous and accurate iron abundances decreases the spread 
along the radial gradient (G13). This means that they can be adopted 
to investigate the fine structure of the metallicity distribution 
\citep{Lepine2011} and the possible occurrence of gaps and/or of 
changes in the slope \citep{Lepine2013}.

iii) Ages -- The central helium burning phases of intermediate--mass 
($M\sim$3.5--10 $M_\odot$) stars take place along the so--called 
blue loops. During these phases an increase in stellar masses causes 
a steady increase in the mean luminosity. These are the reasons why 
classical Cepheids do obey to a Period-Age relation. However, the 
ages covered by Cepheids is quite limited ($\approx$20--400 Myr).     
Most of the current chemical evolution models do predict a steady 
decrease in slope of the metallicity gradient as a function of 
age \citep{Portinari2000,Cescutti2007,Minchev2013}. 
However, we still lack firm estimates of this effect 
since homogeneous estimates of distance, age and chemical 
composition for a large sample of intermediate and old open 
clusters \citep{Salaris2004,Carraro2007b,Yong2012} are still 
missing.

In this investigation we provide new accurate and homogeneous 
iron abundance estimates  for 42 Galactic Cepheids based on 
high spectral resolution, high signal-to-noise ratio (\snr) 
spectra acquired with UVES at VLT.  The total sample includes estimates 
for 75 Cepheids (74 Classical Cepheids and one Type II Cepheid --DD Vel-- 
that will be discussed in a forthcoming paper), whose abundances 
were partially published in \cite{Genovali2013}. 
Moreover, we also analyzed three high spectral resolution spectra 
for the Cepheids --TV CMa, RU Sct, X Sct-- collected with NARVAL at 
the T\'elescope Bernard Lyot (TBL)\footnote{Based on observations collected 
with TBL (USR5026) operated by OMP \& INSU under programme 
ID L072N06 (PI: B. Lemasle).} that we adopted to double check 
current iron abundance estimates. We also added 
a new estimate of the FEROS spectrum for the Cepheid CE Pup whose 
metallicity was already available in the literature \citep{Luck2011a}.  

The above iron abundances were complemented with similar estimates 
provided either by 
our group \citep[][53 Cepheids]{Lemasle2007,Lemasle2008,Romaniello2008} 
or available in the literature \citep[][322 Cepheids]{Yong2006,Luck2011a,Luck2011b}.
We ended up with a sample of 
450 Classical Cepheids i.e. the 73\% of the entire sample of known Galactic Cepheids 
according to the Classical Cepheids list in the GCVS (candidate Cepheids 
are excluded from this estimate).  
For the entire sample, we estimated homogeneous distances based on reddening-free 
near infrared Period--Wesenheit relations  \citep{Inno2013}. 

This is the eighth paper of a series devoted to chemical composition of Galactic
and Magellanic Cepheids (see the reference list). The name of the project is
DIsk Optical and Near infrared Young Stellar Object Spectroscopy (DIONYSOS).
The structure of the paper is the following. In \S 2 we present the
spectroscopic data sets adopted in the current investigation and the
method adopted to estimate the iron abundances. The photometric data and
the individual distances are discussed in \S 3, together with a detailed
analysis of the errors affecting Cepheid distances. \S 4 deals with the
metallicity gradient, while in \S 5 we investigate the dependence of the
metallicity gradient on stellar age. In this section the metallicity
gradient is compared with the metallicity gradients based on younger
tracers (OB stars) and with intermediate-age tracers (open clusters).
In \S 6 we address in detail the fine structure of the metallicity
gradient and perform a thorough analysis of the Cepheid radial
distribution across the Galactic disk.
\S 7 deals with the longstanding open problem concerning the dependence
of the luminosity amplitude on the metallicity, while in \S 8     
we summarize the results and briefly outline the future developments
of this project.


\section{Spectroscopic data and iron abundances}\label{sec:metallicity}
\subsection{Spectroscopic data}
\begin{figure*}
 \centering
 \includegraphics[width=16cm]{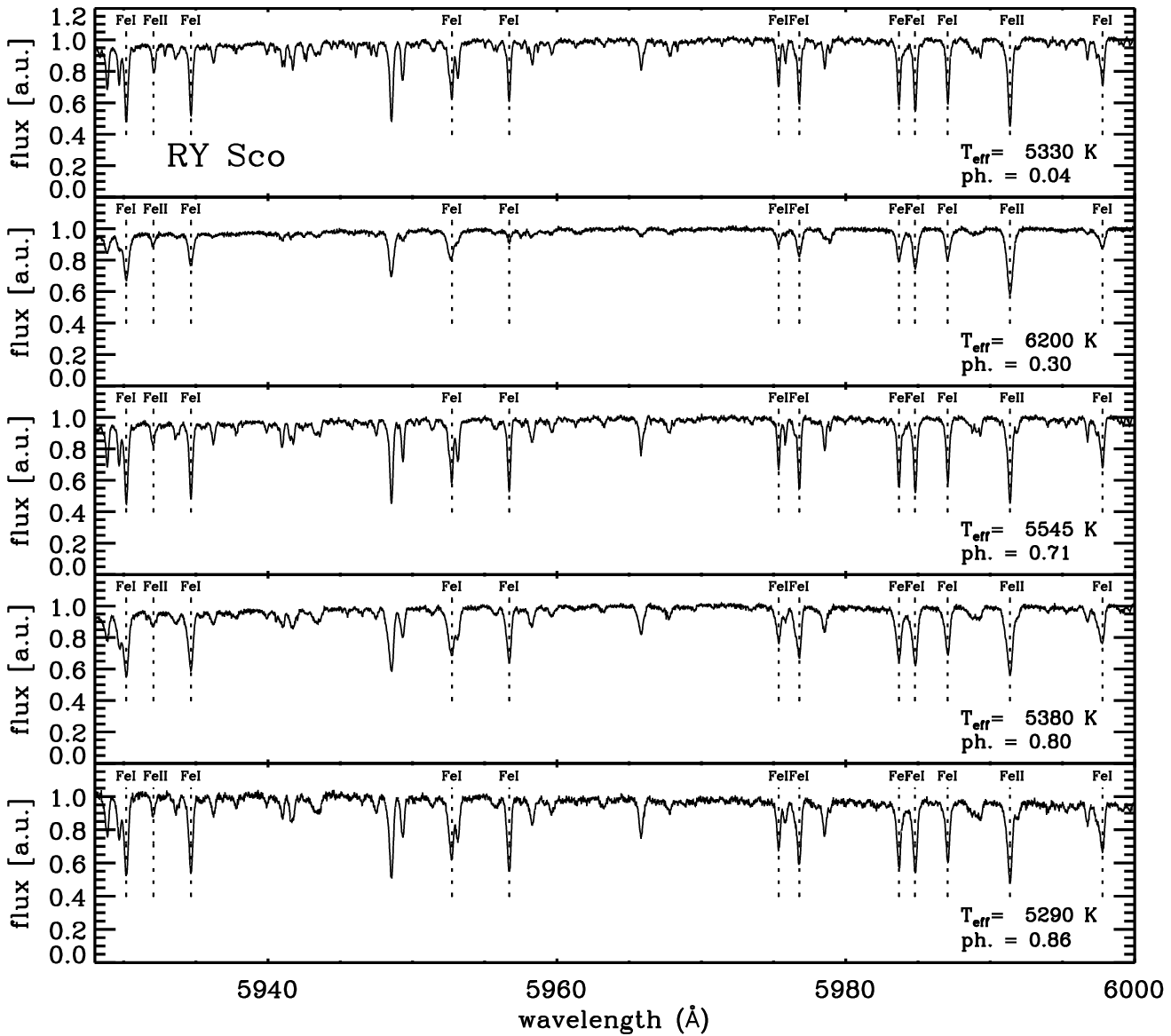}
 \caption{High-resolution (R$\sim$38,000) UVES spectra of solar metallicity Galactic 
 Cepheid RY Sco ([Fe/H] = 0.01 $\pm$ 0.06). From top to bottom the spectra 
 were plotted for increasing pulsational phase. (see the corresponding 
 \teff \ curve in Fig.~\ref{fig:RYSco}).
 The SNR of this sample is around 100 in the spectral range $\lambda$ $\sim$ 5650--7500 \AA. 
 The vertical dashed lines display selected \fei \ 
 ($\lambda\lambda$5930.17, 5934.66, 5952.73, 5956.7, 5975.35, 5976.78, 
 5983.69, 5984.79, 5987.05, 5997.78 \AA) and \feii \ 
 ($\lambda\lambda$5932.06, 5991.37 \AA) lines included in our 
 abundance analysis \citep[see Table~1 and][]{Romaniello2008}. \label{fig:fig1}}
 \end{figure*}

In this investigation we present a spectral analysis based on high-resolution 
(R$\sim$38,000) spectra collected with the UVES spectrograph available 
at the Nasmyth B focus of UT2/VLT Cerro Paranal telescope.
Multi--epoch spectra for eleven Galactic Cepheids were collected during 
observing run B (P89)
. This sample includes 
44 high-resolution 
spectra (from four to six spectra per star) for a total of eleven Cepheids.
The covered spectral range is 4726--5804 \AA \ and 5762--6835 \AA \ over the 
two chips, collected by only using the red arm configuration and the cross 
disperser CD\#3 (central wavelength at 580 nm). The \snr \ ranges from 
$\sim$50 to $\sim$300 (see Fig.~\ref{fig:fig1}). The seeing during the 
observations was ranging from 0.5 to 2.5 arcsec, with a typical mean 
value of 1.2 arcsec, while the exposure times changed from 20 to 1400 sec.

We make use of an additional UVES sample
partially presented in G13.
The spectra were collected at random pulsational phases between 
2008 October and 2009 April using the DIC2 (437+760) configuration  
with blue and red arms operating in parallel. The two arms 
cover the wavelength intervals $\sim$3750--5000 \AA \ and 
$\sim$5650--7600/7660--9460 \AA \ (two chips in the red arm). 
The exposure time ranges from  80 to 2000 sec, while the seeing 
ranges from 0.6 to 2 arcsec with a mean value of 1.2 arcsec. 
The \snr \ is typically better than $\sim$100 for all the echelle 
orders. The complete spectroscopic sample includes 84 spectra for 
a total of 77 Cepheids. The spectra of three Cepheids 
--WW Mon, V641 Cen, GQ Ori-- were analyzed, but they were not 
included in the abundance analysis because the SNR ratio of the 
individual spectra was not good enough. 
The entire sample of UVES spectra were reduced using the ESO UVES 
dedicated pipeline REFLEX v2.1 \citep{Ballester2011}.

For three Cepheids --TV CMa, RU Sct, X Sct-- we also analyzed high \snr \ 
spectra collected with NARVAL at TBL. NARVAL has a spectral resolution 
of $\sim$75,000 and covers the wavelength range 3700-10500 \AA. 
The NARVAL spectra were reduced using the data reduction software 
Libre-ESpRIT, written by 
Donati\footnote{http://www.cfht.hawaii.edu/Instruments/Spectroscopy/Espadons/Espadons\_esprit.html}. 
We also included the re-analysis of a FEROS\footnote{Pre-reduced 
spectra are available at http://archive.eso.org/wdb/wdb/eso/repro/form} 
spectrum for CE Pup \citep{Luck2011a} to better constrain possible 
systematic differences in the metallicity estimates of the outer disk.

As a whole, we provide in this investigation an updated spectroscopic 
estimate of the iron abundance for 42 Classical Cepheids located either 
in the outer disk or in the solar neighbourhood. Together with the iron 
abundances provided by G13 we ended up with a homogeneous metallicity 
sample for 74 Classical Cepheids.

\subsection{Method}\label{sec:method}

\begin{figure*}
 \centering
 \includegraphics[width=17cm]{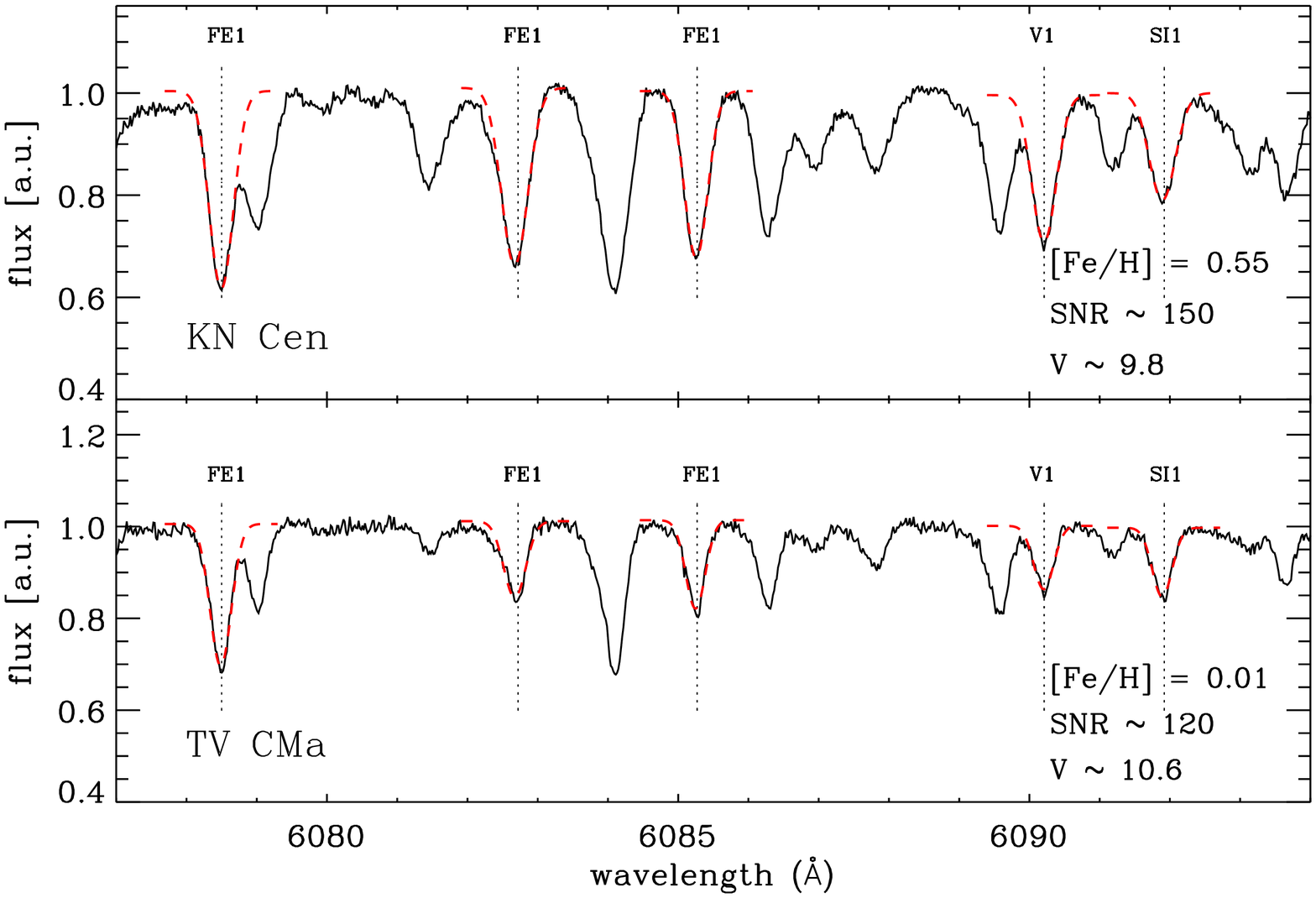}
 \caption{High-resolution (R$\sim$38,000) UVES spectrum of 
 KN Cen ([Fe/H] $=0.55\pm0.12$, \citealt{Genovali2013}) and 
 TV CMa ([Fe/H] $=0.01\pm0.07$, \citealt{Genovali2013}).
 The apparent visual magnitude and the SNR in the spectral range 
 $\lambda$ $\sim$ 5650 - 7500 \AA \ (red arm) are also labeled. 
 The vertical dashed lines display selected lines  
 (\fei \ 6078.50, \fei \  6082.72, \fei \ 6085.27, 
 \vi \ 6090.21, \sii \ 6091.92 \AA) adopted to estimate the 
 individual \teff \ with the LDR method \citep{Kovtyukh2000} 
 (red dashed lines). \label{fig:LDR}}
 \end{figure*}

We implemented a dedicated semi-automatic procedure able to determine 
the continuum and to fit the line profile by single or double Gaussian 
functions (see Fig.~\ref{fig:LDR}), depending on the line 
blending. We adopted the iron linelist presented in \cite{Genovali2013} 
and typically we measured the equivalent width (EW) of $\sim$ 100 - 200 \fei 
\ and $\sim$ 20 - 40 \feii \ lines, depending on the specific spectral range.

The abundance determination was performed by using the code {\it calrai} originally  
developed by \cite{Spite1967} and continuously updated since then. Once fixed the 
atmospheric parameters, the code performs an interpolation over a grid of LTE, 
plane-parallel atmosphere models (MARCS, Gustafsson et al, 2008) and provides 
[Fe/H] and its intrinsic error. The abundances of the other elements will 
be discussed in a forthcoming paper.

For each spectrum we computed the curves of growth for both neutral and 
ionized iron. The process is iterated until a good match between the 
predicted and observed equivalent widths (and thus the curves of growth) 
is obtained.

The effective temperature --$T_{eff}$-- for individual spectra was 
independently estimated by using the line depth ratio (LDR) method 
(Kovtyukh \& Gorlova 2000). Typically, we measured two dozen of 
LDRs per spectrum (see e.g. Fig.~2). The estimated $T_{eff}$ was validated 
by ensuring that $\fei$ does not depend on the excitation potential 
($\chi_{ex}$) i. e. the slope of \fei vs $\chi_{ex}$ is minimal.
The surface gravity --log g-- was derived by imposing the ionization 
balance between \fei and \feii. The micro-turbulent velocity --vt-- 
was estimated by minimizing the $\fei$ vs EW slope. The atmospheric 
parameters estimated for each spectrum are given in Table~\ref{tab:table_atm_par}.

The maximum EW value included in the analysis varies 
according to the metallicity itself and on the atmospheric parameters 
of the star. For a large fraction of our spectra we were able to 
use only relatively weak lines (EW $\le$ 120 m\AA) located along 
the linear part of the curve of growth. In a few cases the spectra were 
lacking of a significant number of weak lines (less than two \feii \ lines), 
therefore, we included in the analysis also lines with EW up to 180 m\AA. 
The latter ones cause a mild increase in the uncertainties affecting the 
correlated atmospheric parameters and become of the order of 
$\Delta$\logg \ 0.3 dex and $\Delta$\vt \ 0.5 km/s. The impact that typical  
uncertainties on effective temperature, surface gravity and microturbulent 
velocity have on the mean iron abundance are listed in Table~\ref{tab:error}.   
Data given in this table indicate that the difference in iron is on average 
smaller than 0.2 dex. Moreover, the difference in iron does not seem to 
depend, within the uncertainties, on the pulsation phase.  

The mean iron abundances given in column 8 of Table~\ref{tab:tab_distances} 
are the weighted mean of \feih \ and \feiih \ with associated errors, i.e. 
$\sigma_{Fe}$=$\sqrt{\sigma_{\fei}^2+\sigma_{\feii}^2}$, 
where $\sigma_{\fei}$ and $\sigma_{\feii}$ are the standard deviations 
of \feih \ and \feiih \ estimates given by the lines measured in a single spectrum.
For Cepheids in our sample with multiple measurements the weighted average abundance 
and the standard deviation  $\sigma_{Fe}$=$\sqrt{\sum_i \sigma_{Fe,i}^2}$ are 
also listed. 
The iron abundances were estimated by assuming the solar iron abundance 
provided by \cite{Grevesse1996}, i.e. A(Fe)${_\odot}$ = 7.50.

In order to validate current iron abundances we adopted the NARVAL spectra, 
since they have a spectral resolution that is a factor of two larger than 
the UVES spectra and similar signal-to-noise ratios. We found that the 
the iron estimates for X Sct, TV CMa and RU Sct based on the NARVAL spectra 
agree quite well with those based on UVES spectra, and indeed the 
difference is on average smaller than $\sim$0.1 dex.

\section{Photometric data and distance estimates}\label{sec:distances}

\subsection{Photometric data}
In order to provide a homogeneous sample of Galactocentric distances ($\Rg$), 
we adopted near infrared (NIR) photometric data together with the reddening-free 
Period-Wesenheit relations in $J, H, K_s$ bands derived by \cite{Inno2013}. 
We estimated individual distances for a significant fraction of Galactic 
Cepheids (93\% of the known Galactic Cepheids). 
To improve the precision of individual Cepheid distances,  we adopted the 
NIR photometric catalogs provided by \cite{Laney1992} and by \cite{Monson2011}. 
The above subsamples were complemented with NIR photometry from the 2MASS 
catalog.

The SAAO data set includes published mean magnitudes from \cite{Laney1992} 
and new multi-epoch measurements (C.D. Laney, private communication).
The individual NIR measurements of the former sample cover the entire pulsation cycle 
and  the accuracy of mean $J, H, K_s$ magnitudes is typically better than 
0.01 mag. Some of the Cepheids in the latter sample lack a detailed 
coverage of the light curve. For these objects the number of phase 
points ranges from four to 14 and they are marked with a dagger in the 
column notes of Table~\ref{tab:tab_distances}. The mean magnitudes were estimated 
using a cubic spline. The SAAO NIR magnitudes were transformed into the 2MASS 
photometric system by using the transformations provided by \cite{Koen2007}.     

We also adopted the NIR photometric catalog from \cite{Monson2011}. They 
provided accurate NIR magnitudes for 131 northern hemisphere Cepheids. Their 
NIR mean magnitudes were transformed into the 2MASS photometric system using 
the calibrating equations provided by the same authors. The measurements 
properly cover the entire pulsation cycle and the typical accuracy on the 
mean magnitudes is better than 0.01 mag.

The above samples were complemented with 2MASS single-epoch 
NIR observations \citep{2MASS}. The mean magnitude for 
Fundamental (FU) Cepheids was estimated by using single-epoch 
photometry and the light-curve template provided by 
\cite{Soszynski2005}. The pulsation properties required to 
apply the template (epoch of maximum, optical amplitudes, 
periods) come from the General Catalog of Variable 
Stars\footnote{\url{http://www.sai.msu.su/gcvs/gcvs/index.htm}}
\citep[GCVS;][]{Samus2009}, with the exception of few objects 
for which we adopted the pulsation periods provided by \cite{Luck2011b}.
The error on the mean NIR magnitudes was estimated as 
$\sigma_{J, H, K_s}^2 = \sigma_{phot}^2+ \sigma_{temp}^2$, 
where $\sigma_{phot}$ is the intrinsic photometric error -- typically 
of the order of 0.03 mag for the Cepheids in the 2MASS sample --  
and $\sigma_{temp}=0.03$ mag is the uncertainty associated with the 
intrinsic scatter of the template.

We did not estimate the NIR mean magnitudes of classical Cepheids pulsating either 
in the first overtone (FO) or as mixed--mode pulsators (''CEP(B)''). Their 
mean magnitudes are the original single-epoch 2MASS measurements, because 
we still lack either the light-curve template for FOs or the epoch of 
maximum. This subsample is marked with an asterisk in the last column 
of Table~\ref{tab:tab_distances}. 

In order to provide an estimate of the uncertainty affecting distance 
estimates based on NIR single-epoch data (FU and FOs), we associated 
to this photometric sample a cautionary total error of 
$\sigma_{J, H, K_s} = \sqrt{\sigma_{phot}^2+ (A_{J, H, K_s}/2)^2}$, 
where ($A_{J, H, K_s}/2$) is the semi-amplitude in the specific band. 
The NIR amplitudes were estimated by using empirical relations for 
the ratio between optical and NIR amplitudes. 
In particular, we adopted the $A_{NIR}/A_I$ ratios provided by 
\cite{Soszynski2005} for FU Cepheids with $\log P$ $\le$ 1.3:  
$A_J/A_I=0.63$, $A_H/A_I=0.50$, $A_K/A_I=0.49$ mag. 
We estimated the amplitude in the I-band --$A_I$-- by using 
the optical ratios $A_I=0.62 A_V$ and $A_I=0.42 A_B$ according 
to the intrinsic parameters available in the GCVS. 

For the FO pulsators we adopted the ratio between optical and 
NIR amplitudes for FU Cepheids with 
$\log P$ $\le$ 1.2 \cite[see][]{Klagyivik2009}. This assumption 
relies on the theoretical and empirical evidence that FOs, once 
their period is fundamentalized, display pulsation properties 
very similar to FU Cepheids with periods shorter than $\log P$ $\le$ 1.2.  

We compared the above estimates with a dozen of complete NIR FOs 
light-curves available in the Laney's sample and we found that 
in every case the observed ratios are quite similar or lesser 
than the estimated ones. For double-mode and putative classical 
Cepheids we used instead the relations $A(NIR)/A(V)$ provided 
by \cite{Soszynski2005} for classical $\delta$ Cepheids. 

For the objects in common in more than one sample 
(SAAO, \citealt{Monson2011}, 2MASS) we adopted 
the most accurate mean magnitude values.

\subsection{Distance determination}

The individual distance moduli were estimated as the weighted mean of the 
three different distance moduli obtained by adopting the NIR 
(\textit{H, J-H}; \textit{K, J-K}; \textit{K, H-K}) Period-Wesenheit (PW) relations  
provided by Inno et al. (2013). The individual distance moduli are 
listed in column 9 of Tables~\ref{tab:tab_distances} and \ref{tab:tab_distances_all} 
with their uncertainties.
The Galactocentric distances listed in column 10 of Tables~\ref{tab:tab_distances}  
and \ref{tab:tab_distances_all}  were estimated 
assuming a solar Galactocentric distance of 7.94$\pm$0.37$\pm$0.26 kpc 
\citep[][and references therein]{Groenewegen2008,Matsunaga2013}. 
The final error on $\Rg$ accounts for errors affecting both the solar 
Galactocentric and heliocentric distances. 

We tested that differences among individual distances based 
on single-epoch 2MASS photometry with those based either on 
SAAO or on \cite{Monson2011} NIR photometry are marginal 
(3\% on average with a standard deviation of 7\%). 
We also compared current individual distances based on 
NIR PW relation with individual distances estimated using 
two different flavors of the IRSB method and we found that 
the mean difference over the entire sample ranges from 
8 $\pm$ 2\%  (Groenewegen 2013, $\sim$ 130 stars in common) 
to  4 $\pm$ 2\% (Storm et al. 2011a, $\sim$ 80 stars in common).
The mean difference between our distances and the distances 
from \cite{Luck2011b} based on optical Period--Luminosity relations 
is of the order of 3 $\pm$ 1\%  ($\sim$ 400 stars in common). On the 
other hand, the typical dispersion between current and literature 
distances ranges from 17\% (our--Luck) to 22\% (our--Groenewegen). 
Thus suggesting that the 
use of homogeneous NIR photometry and solid distance diagnostics 
have a major impact in the decrease of the intrinsic dispersion 
of Galactocentric distances.

\section{The metallicity gradient}\label{sec:gradient}

\subsection{Spectroscopic data sets}\label{sec:abundances}
We compared our new homogeneous estimates 
(current sample plus stars in \citealt[G13]{Genovali2013}) with the iron 
abundances provided either by our group (\citealt[LEM]{Lemasle2007,Lemasle2008}; 
\citealt[ROM]{Romaniello2008}; \citealt[PED]{Pedicelli2010}) or in 
literature (\citealt[LII]{Luck2011a}; \citealt[LIII]{Luck2011b}; 
\citealt[SZI]{Sziladi2007}; \citealt[YON]{Yong2006}).

The increase in the number of Cepheids in common among the different 
data sets allowed us to better evaluate the possible occurrence of 
a systematic difference in the metallicity distribution.
The difference in iron abundance among the different samples are 
the following: 

$\Delta [Fe/H]$(LIII-ROM)$=0.11\pm 0.11$ (22),

$\Delta [Fe/H]$(LII-LEM)$=0.08\pm0.12$ (51), 

$\Delta [Fe/H]$(LIII-YON)$=0.34\pm0.20$ (20), 

$\Delta [Fe/H]$(LII-G13)$=-0.05\pm0.11$ (45),
   
$\Delta [Fe/H]$(LIII-G13)$=0.03\pm0.08$ (33). 

The numbers in parentheses give the number of objects in common among the 
different data sets. The difference with the double-mode Cepheids by 
\cite{Sziladi2007} was not estimated, since we only have one object in common.

The typical difference is on average smaller than 0.1 dex. Our results for 
CE Pup and HW Pup further support the systematic difference between iron 
abundances provided by \citet[]{Yong2006} and similar estimates available in the 
literature \citep{Lemasle2008,Luck2011a,Luck2011b}.
In order to provide a homogeneous metallicity scale for a large sample 
of Galactic Cepheids, we applied the above differences to the quoted 
data sets. The column 7 of Table~\ref{tab:tab_distances_all} lists 
the original iron abundances, while the column 8 gives the re-scaled 
iron abundance.

\subsection{The iron abundance gradient}
In this investigation we analyze the metallicity gradient using 63 homogeneous 
metallicity estimates based on single-epoch UVES spectra. Among them 33 
were already presented in \cite{Genovali2013}. We also include in the 
analysis new weighted mean abundances for eleven Cepheids 
observed from four to six times with UVES at random pulsational phases 
(V340 Ara, AV Sgr, VY Sgr, UZ Sct, Z Sct, V367 Sct, WZ Sgr, XX Sgr, 
KQ Sco, RY Sco, V500 Sco), collected either in observing run A (P81--P82, 
with the exception of V500 Sco) or in observing run B (P89, see 
Table~\ref{tab:table_atm_par}). We confirm the previous findings by 
\cite{Andrievsky2005} and references therein 
that the iron abundance estimates, within the errors, are not phase-dependent.
Moreover, we provide an independent estimate of the FEROS spectrum of the outer 
disk Cepheid CE Pup whose iron abundance was originally determined by 
\cite{Luck2011a}. A more detailed analysis of the multi-epoch spectra 
will be presented in a forthcoming paper.

The top panel of Fig.~\ref{fig:grad} shows the iron abundances based on
UVES multi-epoch spectra of the observing run B (eleven, dark red circles),
on UVES single-epoch spectra of the observing run A (30, red circles) and
on the FEROS spectrum (light blue circles) as a function of the Galactocentric
distances ($\Rg$).
The blue circles display the iron abundances provided by \cite{Genovali2013}
based on UVES single-epoch spectra of the observing run A (33 stars).
Together with the current sample, we also included iron abundances for Galactic
Cepheids estimated by our group using the same approach and similar data
(\citealt{Lemasle2007,Lemasle2008}; 39 objects, green circles;
\citealt{Romaniello2008}; 14 objects, yellow circles).
Current data set covers a range in Galactocentric distances of more than 10 kpc
(4 $\lesssim R_G \lesssim R_G$ 15 kpc). We estimated the metallicity gradient
(dashed line) and we found [Fe/H]=0.49$\pm$0.03 - 0.051$\pm$0.003 $R_G$/kpc.
The new slope and zero--point agree quite well with similar estimates available
in the literature \citep{Luck2011b,Lemasle2013}. The spread in iron appears
to be homogeneous over the entire galactocentric range, but in the innermost
disk regions it increases and becomes of the order of 0.5 dex.

We also included Cepheid iron abundances available in the literature:
  
\cite{Yong2006,Sziladi2007,Luck2011b,Luck2011a} (322 objects, black circles).
The priority for objects in common among different data sets was given
to the current sample, then to iron abundances obtained by our group
and finally to abundances available in the literature.

It is worth mentioning that we have been able, thanks to the current 
large and homogeneous data set of NIR mean magnitudes, to include in 
the analysis of the metallicity gradient 18 Cepheids for which the 
iron abundance was provided by \cite{Luck2011b}, but for which the 
individual distances were not available. 
We ended up with a sample of 450 Cepheids with a homogeneous 
metallicity scale and a homogeneous distance scale.

The metallicity gradient we found is
[Fe/H]= 0.57$\pm$0.02 -0.060$\pm$0.002 $R_G$/kpc, in very good agreement
with previous results from \cite{Luck2011b} based on a similar number of
Cepheids. Note that to identify possible
outliers, we performed a preliminary gradient estimate and we found
[Fe/H]= 0.54$\pm$0.02 -0.057$\pm$0.002 $R_G$/kpc.
Subsequently, we neglected four Cepheids --BC Aql, HK Cas, EK Del, GP Per--
with a gradient residual greater than 3$\sigma$. Three out of the four
neglected Cepheids are classified in the GCVS as candidate Cepheids (CEP),
while HK Cas is very high on the Galactic plane and it has been classified
as an Anomalous Cepheid by \cite{Luck2011b}.

The new metallicity gradient still shows a large intrinsic dispersion
around the Solar Circle and in the outer disk (Fig.~\ref{fig:grad}) with
the possible occurrence either of a change in the slope or of a shoulder
for $7\le\Rg\le$10 kpc, as suggested by
\cite{Twarog1997,Caputo2001,Andrievsky2004}.

To further constrain the nature of the spread in iron along the
metallicity gradient the anonymous referee suggested to check its
depdendency on the distance from the Galactic plane. We selected
the Cepheids in our sample with a distance above the Galactic plane
smaller than 300 pc and we found that the gradient is quite similar:
[Fe/H]=0.49$\pm$0.03 - 0.052$\pm$0.004 $R_G$/kpc. We performed the
same test by using the entire sample and the gradient is once again
minimally affected, and indeed we found
[Fe/H]= 0.53$\pm$0.02 -0.055$\pm$0.002 $R_G$/kpc.
The subsample located closer to the Galactic plane was also adopted
to constrain the spread in iron of the outer disk ($R_G \ge13$ kpc).
We found that the spread decreases from 0.17 dex (30 Cepheids) to
0.13 dex (9 Cepheids). This means that the difference decreases   
from 13\% to 5\% higher than the mean spread over the entire 
disk (0.11 dex).

The referee also noted that the spread in iron abundance around the
Solar circle is larger than the spread in the region between 10 and
14 kpc and suggested that the difference might be caused by a different
azimuthal distributions of the Cepheids in the two disk regions.
To further constrain the dependence of the spread on the azimuthal
distribution we estimated, following Genovali et al. (2013), the
metallicity distribution of the four Galactic quadrants. We found that
the $\sigma$ of the four distributions attain very similar values
(0.013$\pm$0.01 dex), while the mean iron abundance increases by
almost 0.2 dex when moving from the quadrants I/II to the quadrants
III/IV (see Fig.~3 in Genovali et al. 2013).
The reader interested in a more quantitative analysis of the variation
of the spread along the metallicity gradient is referred to section 6.

The iron abundances of the current investigation cover the outer disk 
$\Rg\ge$ 13 kpc and together with iron abundances provided either by our 
group or available in the literature do provide a detailed sampling over 
a broad range of Galactocentric distances ($4 \le \Rg \le 19 $ kpc). 
Data plotted in Fig.~\ref{fig:grad} display a steady increase in the metallicity 
dispersion when moving from the solar circle to the outer disk.
It is interesting to note that Cepheids located in the outer disk 
also show larger distances from the Galactic plane ($|z|$ $\ge$  400 pc) when 
compared with inner disk and solar circle Cepheids 
(see Fig.~\ref{fig:cepazim}). However, we did not find a clear correlation 
between distance from the Galactic plane and metallicity. 
To constrain on a more quantitative basis the above difference, 
we analyzed the Cepheid azimuthal distribution and we found, 
as expected \citep{Kraft1963}, 
that they are on average $\sim$38$\pm$13 pc below the Galactic plane 
and their $\sigma$ is $\sim$270 pc. However, the fraction of 
Cepheids located at distances from the Galactic plane larger than 
1$\sigma$ increases from 4\% for $\Rg$ smaller than 10 kpc to 38\% 
at larger Galactocentric distances.      

It is worth mentioning that the evidence of an increase in the 
dispersion of the iron abundance in the outer disk is further supported by 
the fact that the use of homogeneous iron abundances has further decreased 
the intrinsic spread from 0.12 (see Fig.~2 of \citealt{Genovali2013}) to 
0.10 dex (see Fig.~\ref{fig:grad}) for $\Rg$ $\sim$ 11 -15 kpc.

\section{Comparison between Cepheid and independent metallicity gradients}

\subsection{Young tracers}

During the last few years several investigations have addressed the open 
problem concerning the age dependence of the metallicity gradient. This 
issue has been investigated not only from the empirical 
\citep[see, e.g.,][]{Maciel2003,Nordstrom2004,Henry2010,Yong2012} 
but also from the theoretical point of view. In particular, it has been 
discussed the role that different stellar tracers can play in constraining 
the chemical tagging not only in spatial distribution but also in time 
(Freeman \& Bland-Hawthorn 2002).   

To further constrain the age effect on the metallicity gradient we 
collected several abundance gradients based on different stellar 
tracers. 

Fig.~\ref{fig:plotHII_OBs} shows the comparison between
the metallicity gradient based on Cepheids with the iron abundance
of almost three dozen of early B--type stars (red triangles) located
either in the solar neighborhood \citep{Nieva2012} or in the
nearby Orion star forming region \citep{Nieva2011}. The
key advantage of this set of measurements is that they are based
on high--resolution, high signal--to--noise spectra, they are
homogenous and they also account for non=LTE effects \citep{Przybilla2011}.

The comparison of the iron abundances is further supporting the evidence
that early B--type stars and classical Cepheids display similar abundance
in the solar neighborhood. The spread in iron of the B--type stars is
smaller (see the red vertical error bar plotted in the right corner)
compared with the Cepheids, but they also cover a narrow disk region.
Note that the comparison appears even more compelling if we account
that B--type stars are the typical progenitors of classical Cepheids.

The above scenario concerning the iron abundance gradient of young stellar 
tracers shows a stark difference when compared with iron abundances of 
young stars (red supergiants, luminous blue variables, Wolf--Rayet and 
O-type stars) located either in the Nuclear Bulge or in the near end of 
the Galactic Bar \citep{Martins2008,Davies2009a,Davies2009b,Najarro2009}.
Indeed, the above spectroscopic measurements  
suggest either a solar or a subsolar iron abundance. This finding has been 
soundly confirmed by \cite{Origlia2013} by using high-spectral resolution 
($R\sim$50,000) NIR spectra collected with GIANO at TNG. They found that 
the mean iron abundance of three RSGs located in the RSGC2 cluster are 
sub-solar. This finding does not seem to be supported by recent chemical 
evolution models by \cite{Minchev2013}, since they predict in the 
innermost Galactic regions present days super-solar iron abundances.    

The open clusters (OCs) have several advantages as stellar tracers of the 
Galactic thin disk. i) They typically host a sizeable sample of RGs, this 
means that multi-object spectrograph can provide very accurate measurements 
for both iron and $\alpha$-elements. ii) Their distances can be evaluated 
with good precision by using the main sequence fitting. iii) They trace 
a significant fraction of the Galactic disk (see the WEBDA 
website\footnote{http://webda.physics.muni.cz/}) and their 
ages range from several hundred of Myrs to several Gyrs. The main drawback 
is that they are affected by high reddening and quite often by 
differential reddening.  
To fully exploit the advantages in using OCs to constrain the metallicity 
gradient, we selected a sample of 67 OCs for which are available in the 
literature spectroscopic measurements of iron abundances. To provide a 
homogeneous metallicity scale for OCs, the individual estimates were 
rescaled to the solar iron abundance we adopted in this 
investigation. For the OCs with multiple estimates of the iron abundance 
in the literature, we typically adopted the most recent measurement.
The columns 4 and 5 of Table~\ref{tab:OCs} give both the original and 
the rescaled iron abundance\footnote{Note that in a few cases we have not 
been able to rescale the iron abundance, since the authors did not 
quote the adopted solar iron abundance.}, while columns 6 and 7 give 
the reference for the metallicity and for the age/distance.

In dealing with Galactocentric distances of OCs, the main source 
of uncertainty is the calibration of the adopted distance indicator. 

Moreover and even more importantly, age estimates are tightly correlated
with the adopted cluster distance, reddening and metallicity. The cluster
age also depends on the evolutionary framework (overshooting, mass loss,
rotation, microphysics) adopted to compute evolutionary models and
cluster isochrones \citep{Bono2001e,Salaris2008,Prada2012,Neilson2012b,Anderson2013}.
To overcome this thorny problem and to limit their intrinsic dispersion
in the metallicity gradient we decided to only use OC with homogeneous
estimates of the four crcuail parameters: age, distance, reddening,
abundance, theoretical framework. In particular, we selected
determinations provided by
\cite{Salaris2004} and \cite{Friel1995} (30 OCs), 
by the Carraro's group (13 OCs),  
by BOCCE\footnote{http://www.bo.astro.it/~angela/bocce.html} (8 OCs),
and by \cite{Friel1993} (8 OCs). We adopted the \cite{Yong2012} values for 2 remaining OCs.
For seven OCs selected by \cite{Cheng2012} we 
adopted the parameters given by WEBDA and for PWM4 the estimates provided by 
\cite{Yong2012}. The Galactocentric distances are listed in column 3 
of Table~\ref{tab:OCs} and they were calculated by using the same value 
of the Sun Galactocentric distance ($\Rg=7.94$ kpc).

The top panel of Fig.~\ref{fig:plot_OCs} shows the comparison 
between Cepheids (black dots) and 44 OCs younger than 3 Gyrs. 
Diamonds display the position of OCs and different 
data sets are marked with different colors (see also Table~\ref{tab:OCs}). 
Data plotted in this panel show that Cepheids and OCs younger than 3 Gyrs 
are characterized by similar trends when moving from the inner to the 
outer disk. The same outcome applies, within the errors, to the intrinsic 
dispersion. 

We found a metallicity gradient for the young OCs in our sample
of [Fe/H]= 0.47$\pm$0.10 - 0.051$\pm$0.010 $R_G$/kpc (see the top panel              
of Fig.~5), in which both the slope and the zero--point attain values
very similar to the Cepheid metallicity gradient (see Fig.~4).

\subsection{Intermediate--age tracers}

The bottom panel of Fig.~\ref{fig:plot_OCs} shows the same comparison as the 
top panel, but for OCs (23) with ages ranging from 3.6 to 9 Gyrs. 
Data plotted in this panel show two distinctive features.  
i) Old OCs display a clear flattening in iron abundance for $R_G\ge15$ kpc. 
ii) The old OCs for Galactocentric distances between the solar circle and 12 kpc  
seem to show a dichotomic distribution. The difference is of the order of several 
tenths of dex, i.e. larger than possible uncertainties affecting individual iron 
abundances. We also checked the position of the seven OCs that are, at fixed 
Galactocentric distance more metal-poor and we found that five out of the seven 
cover a very narrow range in Galactic latitude (y$\sim$3.5 kpc). 
Data plotted in the above figure support the evidence that the metallicity 
gradient depends on age for ages older than $\sim$3 Gyrs. We could also speculate 
that there is a dozen of OCs distributed along a metal--poor plateau with an almost 
constant iron abundance ([Fe/H]$\sim$-0.4) and Galactocentric distances ranging 
from 9 to 21 kpc. 

We estimated the metallicity gradient of the older OCs and we found
of [Fe/H]=0.21$\pm$0.11 - 0.034$\pm$0.009 $R_G$/kpc (see the bottom panel              
of Fig.~5). The slope and the zero--point are significantly shallower
than for Cepheids and younger open clusters and agree quite well similar
estimates avaialble in the literature for old OCs \citep{Carraro2007b,Yong2012}. 
However, our sample of OCs covers a range in age of five Gyrs and the sample 
is modest. More solid constraints call for larger samples of OCs and a wider 
disk coverage.

However, data plotted in the bottom panel of Fig.~\ref{fig:plot_OCs} 
show that the Cepheid iron abundances in the inner disc are more 
metal--rich than thin disk old OCs. Moreover, the Cepheids display a well-defined 
iron gradient when moving from the inner to the outer disc (5 $\le R_G \le18$~kpc).
However, the above evidence relies on stellar populations with significantly
different ages. The Cepheids and the supergiants of the Nuclear Bulge and
of the Bar have ages ranging from a few Myrs to a few hundreds of Myrs.

The above findings indicate that younger tracers appear to be still in {\it situ}, 
i.e. in the same regions where they formed, while the intermediate-age tracers 
appear to be affected both by radial gas flows and by radial migration, as 
suggested by chemical evolution models \citep{Portinari2000,Curir2012,Minchev2013}.
However, current data do not allow us to constrain the timescale within 
whom the metallicity gradient becomes shallower.

\section{The fine structure of the metallicity gradient}

We are facing the evidence that the intrinsic dispersion of the iron metallicity 
gradient is, at fixed Galactocentric distance, systematically larger than the 
expected standard deviation (see the error bar in the right corner of the 
bottom panel of Fig.~\ref{fig:grad}). This circumstantial evidence stimulated 
several investigations aimed at constraining the physical reasons for such a 
broad distribution.  
On the basis of a large Cepheid data se,t \cite{Luck2006} suggested that 
the large dispersion in iron abundance for Galactocentric distances of 
$R_G$ $\sim$ 9-11 kpc was caused by a metallicity island located at 
$l=130^{\circ}$. However, the detection of well defined region 
characterized by a higher iron abundance was not supported in a 
subsequent analysis by \cite{Luck2011b} by using a larger Cepheid 
sample. The evidence of a clumpy metallicity distribution across the 
Galactic disk was also brought forward by \cite{Lemasle2008} and 
by \cite{Pedicelli2009} by using similar samples of Galactic 
Cepheids. However, the evidence was partially hampered by the sample
size and by the lack of a homogeneous metallicity scale.  

\subsection{Identification of Cepheid Groups}

To further constrain the above preliminary evidence we decided to follow
a different approach. We performed a new search for Cepheids Groups 
(CGs) across the Galactic disk. The search in 3D space follows a method 
originally suggested by \cite{Battinelli1991} and applied to Galactic 
Cepheids by Ivanov (2008, hereinafter I08). He adopted $J$,$H$,$K$ 2MASS 
photometry for 345 Galactic Cepheids and identified 18 CGs. Current 
approach when compared with I08 has several advantages: 
i) we are dealing with a Cepheid sample that is 30\% larger and they 
cover more than 15 kpc across the disk;   
ii) individual Cepheid distances are independent of reddening corrections; 
iii) the 56\% of NIR Cepheid mean magnitudes are based on multi-epoch 
light curves.   

We ranked the entire Cepheid sample and arbitrarily selected the first one 
as a pivot and estimated its closest neighbourhood by using their rectangular 
coordinates x,y and z. Then, we selected the second Cepheid as a pivot, but 
we removed from the list the first one. In the next step, we selected the 
third Cepheid in our list as a pivot, but we removed from the list the first 
and the second one. This process is iterated until we rich the last but one 
Cepheid in our list. Thus we are left with a set of N-1 pair distances 
that provide, by definition, a path connecting the entire Cepheid sample, 
the so called "Path Linkage Criterion" \citep{Battinelli1991}.
The two main positive features of the above algorithm are that 
a region with a high concentration of short pair distances 
is also a region in a 3D space with a high concentration of Cepheids.    
Moreover, the use of relative distances on a common path provides 
solid detections of filamentary groups \citep{Battinelli1996}. 

Once we have the set of pair distances, we need to define on the basis 
of our Cepheid sample a characteristic distance called "search distance" 
--$d_S$-- that will allow us to identify candidate Cepheid groups. 
In particular, we define a candidate Cepheid group if $m$ Cepheids 
(with $m\ge$6) have a distance $d < d_S$, where $d_S$ is an arbitrary distance
in kpc. We adopted $d_S$ distances ranging from 0.1 to 0.8 kpc with a 
step in distance of 0.05 kpc. Fig.~8 shows the number of independent 
CGs we detected as a function of the searching distance. The distribution 
of candidate CGs we found is similar to the distribution found by I08. 
However, the current peak of the distribution is slightly smaller (13 vs 18) 
and takes place at smaller $d_S$ distances (0.25 vs 0.40 kpc). Moreover, 
the number of candidate CGs decreases quite rapidly for $d_S$ distances 
larger than $\sim$0.6 kpc while I08 detected CGs at $d_S$ distances 
larger than 1 kpc (see Fig.~3 in I08). The difference might be explained 
with the difference in sample size and in the adopted Cepheid distances.    

Once we have defined the optimal search distance for our Cepheid sample  
we need to define a criterion to constrain how significant is the density 
of the individual candidate CG when compared with the average stellar 
density of its neighbourhood. In particular, we define a bonafide CG 
only the candidate CGs whose density --$\rho_{CG}$-- is four times larger 
than the density of a spherical layer --$\rho_{sl}$-- centered on the center 
of mass of the CG. The outer radius of the spherical layer --$R_{sl}$--    
was fixed in such a way that the volume of the spherical layer 
is three times larger than the volume of the CG. In particular, 
$V_{sl}$=3$V_{CG}$= $4\pi (R_{sl}- R_{CG} )^3$/3 
where $V_{CG}$ is the volume of the candidate CGs estimated by using a
Montecarlo method an by assuming for each Cepheid in the group a radius
equal to the adopted search distance ($d_S$). 
The radius of the candidate CG was defined as 
$R_{CG}$=1.1$\times$ $\delta$, where $\delta$ is the distance of the 
two most distant Cepheids.       
Note that $R_{CG}$ is also by construction the inner radius of the 
spherical layer adopted to estimate the difference in density between 
the candidate CG and its stellar neighbourhood.  

Fig.~\ref{fig:sfere3g} shows a 3D graphical view of the approach we adopted 
to estimate the density of the candidate CGs and the density of the spherical 
layer. The members of the candidate Cepheid group are plotted as magenta 
spheres, while the inner light grey sphere defines $V_{CG}$, i.e. the volume  
of the sphere (with radius equal to $R_{CG}$) adopted to estimate the density 
of the candidate CG ($\rho_{CG}$). 
The dark grey sphere defines $V_{sl}$, i.e. the volume of 
the spherical layer of outer radius $R_{sl}$ and inner radius $R_{CG}$ 
adopted to estimate the average density of the stellar vicinity ($\rho_{rl}$). 

In order to fix the cutoff density for the identification of CGs, we 
performed a series of numerical experiments in which we randomly 
distributed the same number of Cepheids across the Galactic disk and we 
found that their densities are systematically smaller than four times 
the densities of the spherical layers.  By adopting this conservative 
selection criterion we ended up with ten candidate CGs. The overdensities 
of the selected CGs range from 4.1 to 45.7 when compared with their 
stellar neighborhood. The coordinates and the Galactocentric distances 
of the newly identified Cepheids are listed in columns 1 to 5 of 
Table~\ref{tab:islands} together with their diameters and the number of members.
The smaller groups have on average 6--7 members and have sizes of the 
order of half a kpc, while the largest ones have 20--50 members and sizes 
between 1.2 and 1.9 kpc. The above dimensions are similar to the typical 
size of giant molecular clouds (see Fig.~3 in \citealt{Bolatto2008} 
and Table~1 in \citealt{Murray2011}) and to the typical size of giant 
star complexes and superassociations \citep{Elmegreen1983,Efremov1995}. 

To further constrain the spatial distribution of the newly identified 
candidate CGs, the top panel of Fig.~\ref{fig:fig23_island} shows their 
projection (filled circles) onto the Galactic plane. The members of the 
individual candidate CGs are plotted in yellow and confined by ellipses.  
Each group is marked by an increasing Roman number according to the 
Galactocentric radius.
In order to find a correlation between the location of candidate CGs 
and the spiral arms, we also plotted a simplified model of the disk 
spiral structure. We used the logarithmic model presented by 
\cite{Vallee2002} with four arms and a pitch angle of 12 degrees. 
The logarithmic parameter $r_0$ = 2.58 was fixed in such a way that 
the Perseus arm overlaps with the fiducial points of the model 
provided by \cite{Cordes2002}. We adopted this empirical calibration 
because \cite{Xu2013} found that the latter disk model fits quite well 
the parallax data of 30 masers associated with star-forming regions in 
the Perseus and in Sagittarius arms.

The true location of the spiral arms is not well defined, since it 
depends on the adopted tracers whose distance is quite often poorly known. 
However, data plotted in the top panel of Fig.~\ref{fig:fig23_island} 
indicate a correlation between candidate Cepheid groups and star formation 
regions associated with the spiral arms.

\subsection{Residuals of the metallicity gradient}

To further investigate the physical connection of the individual members of 
the candidate CGs, we analyzed the residuals of the metallicity gradient. 
We estimated for each Cepheid in our sample the difference between its 
iron abundance and the iron abundance of the metallicity gradient at 
the same Galactocentric distance.  
To avoid spurious fluctuations in the mean iron abundance, 
we ranked all the Cepheids as a function of the Galactocentric distance 
($R_G$) and estimated the running average by using the first 20 objects
in the list. The mean $R_G$ and the mean residual ($\Delta [Fe/H]$) of 
the bin were estimated as the mean over the individual Galactocentric 
distances and residual abundances of the same 20 objects. We estimated
the same quantities by moving one object in the ranked list until we
accounted for the last 20 Cepheids in the sample with the largest distances. 
The running average is plotted as a black line in the bottom panel 
of Fig.~\ref{fig:fig23_island}. The error on the mean residual for 
individual bins is of the order of a few hundredths of dex. In order 
to provide robust constraints on the possible uncertainties introduced 
by the adopted number of Cepheids per bin and by the number of stepping 
stars, we performed a series of Monte Carlo simulations. The estimated 
mean dispersion of the above simulations is plotted as a vertical 
black line.

Interestingly enough, the residuals display local minima and maxima 
that are significantly larger than the intrinsic dispersion. 
The occurrence of the above chemical inhomogeneities is well defined 
for Galactocentric distances smaller than 11 kpc. Unfortunately, 
the current sample does not allow us to rich firm conclusions 
concerning the outer disk. To constrain the nature of the secondary 
features in the residuals, we overplotted in the bottom panel of 
Fig.~\ref{fig:fig23_island} the position of the ten candidate CGs
(red dots). We adopted the mean Galactocentric distance and the 
mean iron abundance of the individual CGs listed in columns 5 
and 14 of Table~\ref{tab:islands} and subtracted the iron 
abundance of the metallicity gradient at the same $R_G$. The red 
vertical lines display the standard deviation of the iron abundances, 
while the horizontal red lines display the inner and the outer edge 
of the individual CGs (columns 6 and 7 of Table~\ref{tab:islands}).      

Data plotted in this figure show that the residuals in iron abundance 
appear to be tightly correlated with the mean residual abundance abundance 
of the candidate CGs. This finding further supports the evidence that a 
significant fraction of the intrinsic dispersion of the metallicity 
gradient is caused by the presence of Cepheid Groups across the Galactic 
disk with mean metallicities that are either more metal--rich or more 
metal--poor than expected according to a linear mean metallicity gradient. 
The mean periods and their intrinsic dispersions listed in columns 10 
and 11 of Table~\ref{tab:islands} seem to suggest that the candidate CGs 
with negative iron residuals have on average slightly longer periods and 
larger intrinsic dispersions when compared with the candidate CGs showing 
a positive iron residual. However, this evidence could be caused by an 
observational bias, and indeed the former candidate CGs have higher Cepheid 
densities (see columns 9 and 13 of Table~\ref{tab:islands}) when compared 
with the latter ones. The pulsation and evolutionary properties of the 
candidate CGs will be discussed in a forthcoming paper.   

The above empirical evidence have also implications concerning the chemical 
enrichment of the Galactic disk. During the last few years it has been 
suggested that iron and oxygen abundances do show a break in the abundance 
gradient associated with the corotation resonance of the spiral pattern 
\citep[see e.g.][]{Acharova2010}. This evidence applies not only to external 
spiral galaxies \citep{Scarano2011,Scarano2013}, but also to our 
Galaxy \citep{Lepine2011}.  In particular, it has been 
suggested that the breaks in iron, $\alpha$-elements and barium abundance 
gradients of Cepheid and open cluster are caused by the corotation 
resonance located at $R_G\sim$9.0--9.5 kpc \citep{Lepine2013}. 
The occurrence of a discontinuity at the above Galactocentric distance is 
further supported by the presence of a well defined local minimum in the         
Galactic rotation curve (see Figures 1, 3 and 5 in 
\citealt{Sofue2009}). On the basis of the above preliminary 
evidence it has been suggested that the Galaxy is experiencing a bimodal 
chemical evolution, since the kinematic of the gas has opposite directions 
at the corotation resonance of the spiral pattern. 

However, current iron abundances for Galactic Cepheids do not show either a 
break or a jump or a change in the slope for $R_G \sim$9.0--9.5 kpc. Moreover 
and even more importantly, the positive iron residual located at the above 
Galactocentric distance appears to be associated with the five candidate 
CGs that are located across the Perseus arm (see the solid blue line in 
the top panel of Fig.~10). Note that the association between 
the candidate CGs and the Perseus arm does requires a more detailed 
analysis, since we adopted a qualitative model of the Galactic logarithmic 
spiral arms \citep{Vallee2005}. In passing we note that current finding, 
once confirmed by independent stellar tracers, supports the suggestion 
brought forward by \cite{Sofue2009} and by \cite{Sofue2013} 
that the dip in the rotation curve $R_G \sim$9.0--9.5 is caused by a massive 
ring associated with the Perseus arm.

\section{Luminosity amplitudes and metallicity dependence} 

During the last few years the dependence of the luminosity amplitude 
on metallicity has been investigated both from the theoretical 
\citep{Bono2000b} and the observational \citep{Klagyivik2009,
Pedicelli2010,Klagyivik2013} point of view. 
In particular, \cite{Szabados2012}, by using a large sample 
(327) of Galactic Cepheids with accurate pulsation parameters 
and spectroscopic metal abundances, found evidence that the 
luminosity and the radial velocity amplitudes slightly decreases 
with increasing iron abundance. However, no firm conclusion has been 
reached, since \cite{Pedicelli2010} by using a similar data set 
did not find a clear dependence on metallicity. More recently, 
\cite{Klagyivik2013} by using short--period ($\log P\le 1.02$) 
Cepheids found that the $R_{21}$ and the $R_{31}$ Fourier amplitude 
ratios decrease for increasing iron abundance.       

To further constrain the behavior of this interesting diagnostic 
we took advantage of the current sample of Galactic Cepheids with 
homogeneous spectroscopic abundances. We limited our sample to 
fundamental Cepheids for which are available accurate V-band 
amplitudes\footnote{For a small sample of Cepheids the V-band 
luminosity amplitude was not available in the literature. For 
these objects the V-band amplitude was estimated using the 
amplitude relations $A_V=A_I/0.622$, for log $P < 1.02$, and  
$A_V=A_I/0.606$, for log $P > 1.02$ provided by \cite{Klagyivik2009}.}
and we ended up with a sample of 351 Cepheids. 
However, the period distribution of Galactic Cepheids shows 
a short tail in the long--period range when compared with 
Magellanic Cloud (MC) Cepheids \citep{Gascoigne1974,Bono2010,Inno2013,Genovali2013}.    
To overcome possible biases in constraining the metallicity 
dependence we took advantage of the MC Cepheids for which were  
available spectroscopic iron abundances \citep{Luck1992,Luck1998,Romaniello2008}
 and V-band amplitudes (SMC and LMC field Cepheids: 
OGLE III  [\cite{Soszynski2008}, \cite{Soszynski2010}]; 
ASAS  [\cite{Karczmarek2011}, \cite{Karczmarek2012}]; 
 plus data provided by  
\cite{vanGenderen1983}, \cite{Freedman1985}, \cite{Caldwell1986}, 
\cite{vanGenderen1989}, \cite{Caldwell2001}. 
For cluster Cepheids we adopted data provided by 
\cite{Sebo1995} [NGC 1850] and by \cite{Welch1991,Welch1993} [NGC~1866]). 
  
We ended up with a sample of 58 Large Magellanic Cloud (LMC) 
and 19 Small Magellanic Cloud (LMC) Cepheids. Note that in the first 
sample are also included 19 Cepheids belonging to the cluster 
NGC~1866 and 7 Cepheids belonging to the 
cluster NGC~1850. The metallicities and 
the pulsation parameters for the Magellanic Cepheids are listed 
in Table~\ref{tab:amp}. The top left panel of Fig.~\ref{fig:amp} shows the Bailey 
diagram (V-band luminosity amplitude vs logarithmic period) for the 
entire sample (428 stars). The center of the Hertzsprung progression 
at $\log P\le$ 1.02 is quite evident. The reader interested in a more 
detailed discussion concerning the nature of the Hertzsprung 
progression and its metallicity dependence is refereed to 
\cite{Bono2000a,Bono2000b}. 

In order to constrain the dependence of the luminosity amplitude on the 
metallicity we split the entire sample into metal-poor 
([Fe/H]$\le$0.03, red circles) and metal-rich ([Fe/H]$>$0.03, 
green circles). To avoid spurious fluctuations in the mean $A_V$ 
amplitude, we ranked all the Cepheids as a function of the logarithmic 
period and estimated the running average by using the first 25 objects
in the list. The mean $\log P$ and the mean $A_V$ of the bin were 
estimated as the mean over the individual periods and amplitudes 
of the same 25 objects. We estimated 
the same quantities by moving one object in the ranked list until we 
accounted for the 25 Cepheids with the longest periods. The running 
averages for the metal-poor and the metal-rich samples are plotted 
as red and green lines in the bottom left panel of Fig.~\ref{fig:amp}. The error 
on the mean $A_V$ for individual bins is of the order of a few 
hundredths of mag. In order to provide robust constraints on the 
possible uncertainties introduced by the adopted number of Cepheids 
per bin and by the number of stepping stars, we performed a series 
of Monte Carlo simulations. The estimated mean dispersions of the 
above simulations are plotted as vertical green and red lines. 
We also found that we can 
exclude a metallicity dependence of the two subsamples at the 
98\% confidence level. The two running averages plotted in the 
bottom panel display a strong similarity in the short--period 
($\log P\le$ 1.02) range, but the difference increases at 
longer periods. We performed the same analysis by splitting 
the sample in short- and long--period Cepheids and we found that 
the metallicity dependence in the former group can be excluded at 
the 94\% level, while the in the latter one at the 70\% level.   

To further constrain the dependence of the V-band amplitude on 
metallicity, we performed the same analysis but the entire sample was 
split into: metal--poor ([Fe/H]$\le-0.01$), metal--intermediate 
($-0.01 \le$ [Fe/H] $\le +0.01$) and metal--rich ([Fe/H]$\ge +0.01$).     
The three different subsamples are plotted as red, black and green 
circles in the top right panel of Fig.~\ref{fig:amp}, while the three running 
averages and their mean dispersions are plotted in the bottom right 
panel of the same figure. The outcome concerning the metallicity 
dependence is quite similar to the above analysis. We found that the 
metallicity dependence between the metal-poor and the metal-intermediate 
subsamples can be excluded at the 90\% confidence level, while the 
dependence between the metal-poor and the metal-rich subsample at 
the 80\% level. We also split the sample in short-- and long--period 
and we found that the dependence can be excluded at the 93\% and at 
the 90\% level between metal--poor and metal--intermediate Cepheids 
and at the  90\% and at the 77\% level between metal--poor and 
metal--rich Cepheids.   

The above findings indicate that the luminosity amplitude of Galactic 
and MC Cepheids does not display a solid trend with metal abundance.    
However, current analysis should be cautiously treated for two different 
reasons: i) The Galactic sample is still dominated by Cepheids at solar 
iron abundance, since the sample of more metal-poor Cepheids located 
in the outer disk ($R_G\ge$ kpc) is quite limited; ii) The spectroscopic 
abundances for MC Cepheids are dominated by brighter (long--period) 
Cepheids. Firm conclusions concerning the metallicity dependence 
do require larger samples of spectroscopic abundances to further 
constrain the difference in pulsation amplitude and in period distribution.    

\section{Discussion and conclusions}
We performed accurate new measurements of iron abundances for 42 Galactic
Cepheids using high-resolution, high-\snr \ UVES, NARVAL and FEROS spectra.
The iron abundance, for eleven Cepheids located in the inner disk, is 
based on multi-epoch spectra (from four to six) and their intrinsic 
uncertainty is smaller when compared with other Cepheids at super-solar 
iron content.
Current sample was complemented with Cepheid iron abundances based on
high--resolution spectra provided either by our group
\citep{Lemasle2007,Lemasle2008,Romaniello2008,Genovali2013}
or available in literature \citep{Luck2011a,Luck2011b}. 
We ended up with a sample of 450 Cepheids.
To improve the accuracy on the metallicity distribution across the disk,
we estimated homogeneous and reddening-free distances by using near-infrared
Period--Wesenheit relations for the entire sample.

The main findings of the current iron abundance analysis are given 
in more detail in the following.

\begin{itemize}

\item[$\bullet$]  

We found that the metallicity gradient, based on current 
spectroscopic measurements, is linear with a slope of
-0.051$\pm$0.003 dex/kpc, in agreement with recent studies by
\cite{Luck2011a} and \cite{Luck2011b}. The metallicity gradient 
based both on our and on literature iron abundances shows a 
similar slope: -0.060 $\pm$ 0.002 dex/kpc.    
Current estimates agrees 
quite well with the chemical evolution model for the thin 
disc recently provided by \cite{Minchev2013}. In particular, 
they found that the iron gradient is -0.061 dex/kpc for 
Galactocentric distances ranging from 5 to 12 kpc and 
-0.057 dex/kpc for Galactocentric distances ranging from 
6 to 11 kpc. The predicted slopes become marginally shallower 
if they account for stellar radial migrations.     

\item[$\bullet$]  

We estimated the metallicity gradient by selecting the Cepheids 
in our sample with a distance above the Galactic plane smaller 
than 300 pc and we found that it is, within the errors, quite 
similar: -0.052$\pm$0.004 dex/kpc. The same outcome applies 
to the the gradient based on the entire sample, and indeed 
we found: -0.055$\pm$0.002 dex/kpc.
We also found that the spread in iron in the outer disk
($R_G \ge13$ kpc) decreases by more than a factor of two
(0.13 vs 0.17 dex) if we adopt the subsample located closer
to the Galactic plane.

\item[$\bullet$] We also confirm that classical Cepheids in the 
inner disk ($\Rg$ $\sim$5.5--6.0 kpc), just beyond the position 
of the Galactic Bar corotation resonance \citep{Gerhard2011}, 
attain super-solar (\feh$\sim$0.4) iron abundances. This result supports similar 
findings by G13 and by \cite{Andrievsky2002,Pedicelli2010,Luck2011b}.

There is preliminary evidence that the iron abundance in the innermost
Galactic regions (Nuclear Bulge, Galactic Bar) is more metal--poor
than predicted by chemical evolution models (\cite{Minchev2013}). Indeed 
recent spectroscopic iron abundances of young stars (red supergiants,
luminous blue variables, Wolf--Rayet, O-type stars) indicate either solar
or sub--solar abundances \citep{Davies2009a,Davies2009b,Origlia2013}.
On the other hand, chemical evolution models suggest in 
the same regions iron abundances larger than \feh$\sim$0.8
\citep{Minchev2013}.
The above evidence indicate that objects located inside the 
corotation resonance of the bar experienced a different chemical 
enrichment history when compared with Cepheids located just 
beyond this limit.

\item[$\bullet$]  
The new homogeneous Cepheid metallicity distribution is 
characterized by a smaller intrinsic dispersion when 
compared with similar estimates available in the literature. 
We found evidence of a steady increase in the abundance dispersion 
when moving in the outer disk ($\Rg$ > 14 kpc). Current data 
do no allow us to constrain whether this effect is the aftermath 
of outward stellar migrators as recently suggested by \cite{Minchev2012}
 or the consequence of the infall of the  
Sagittarius dwarf galaxy producing a flared outer disk as suggested 
by \cite{Purcell2011}. 

\item[$\bullet$]  
To investigate the fine structure of the metallicity in the disk,
we searched for Cepheids groups following the approach suggested by
\cite{Ivanov2008}. We found ten candidate Cepheids Groups,
i.e. physical aggregation of stars whose mean residual metallicity agrees 
quite well with the trend of the metallicity residuals as a function of 
the Galactocentric distance. The presence of the CGs appears to be the 
main culprit of the fluctuations in the metallicity residuals and of 
the azimuthal effects on the radial gradient. This suggests that 
members of CGs experienced a very similar chemical enrichment history. 
Most of the CGs are located close to spiral arms (Sagittarius-Carina
and Perseus arms) according to a simple logarithmic spiral model 
provided by  \cite{Vallee2005}. The above findings indicate that 
the occurrence of CGs with sizes ranging from OB association/young 
cluster to star complexes/superassociations appear to be largely 
responsible for the intrinsic spread of the iron metallicity gradient. 
Moreover, the association of the metallicity residuals with candidate CGs supports 
the results by \cite{Sofue2013} concerning the association of a local minimum 
in the Galactic rotational curve at $\Rg\sim$=9.5 kpc with the Perseus arm.

\item[$\bullet$]  
We also found that the mean periods of the Cepheids hosted in candidate CGs 
with negative iron residuals have, on average, slightly longer periods and 
larger intrinsic dispersions when compared with the candidate CGs showing
a positive iron residual. Thus suggesting a common star formation episode 
within each candidate CG.
The evidence of possible abundance inhomogeneities in the Galactic disk 
dates back to \cite[][and references therein]{Efremov1995} who suggested that 
the different star complexes might have different star formation histories 
and different interactions with the intergalactic medium. It is clear that 
the abundance information (iron and $\alpha$--elements) will provide a new 
spin to the analysis of their evolutionary and pulsation properties.

\item[$\bullet$]  
To constrain the impact of age on iron abundance gradient, we 
compared the Cepheid iron gradient with those based on OCs.  
Spectroscopic metallicities and homogeneous distances and age were
collected for OCs spanning a large range in age. The OC gradient based 
on clusters younger than 3 Gyrs agrees quite well with the Cepheid 
gradient. 

\item[$\bullet$]  
The comparison  between Cepheids and OCs older than 3 Gyrs is more 
complex. Indeed, we found that old OCs display a clear flattening 
in iron abundance for $R_G\ge15$ kpc. This result supports similar findings 
available in the literature e.g. \cite{Carraro2007b}, \cite{Bragaglia2008}, 
\cite{Magrini2009,Magrini2010}, \cite{Jacobson2011a,Jacobson2011b}, 
\cite{Yong2012}.
Moreover, old OCs located between the solar circle and $\Rg\sim$12 kpc
seem to show a dichotomic distribution. The difference is of the order 
of several tenths of dex and might be due to a selection bias affecting 
the azimuthal distribution. 
However, the comparison of Cepheids iron abundances with similar abundances 
for old OCs further support the evidence that the metallicity 
gradient does depend on age for ages larger than $\sim$3 Gyrs.

\item[$\bullet$]  
We investigate the possible occurrence of a metallicity effect 
on the pulsational amplitude by using a large sample of fundamental 
Galactic and Magellanic Cepheids \citep{Luck1992,Luck1998,Romaniello2008} 
with accurate iron abundances. The comparison of low, medium, and high 
metallicity subsamples indicate that luminosity amplitudes are, within 
current uncertainties, independent of iron abundance.

\end{itemize}

Classical Cepheids appear to be solid young stellar tracers to constrain
the recent chemical enrichment of the Galactic thin disk. Current sample
of Galactic Cepheids is smaller when compared with similar tracers
(OB stars, HII regions, red clump stars, open clusters). However, their
distances, ages and abundances can be firmly estimated. They are ubiquitous
in young star forming regions and the recent identification of classical
Cepheids both in the Nuclear Bulge and in the Galactic Bar \citep{Matsunaga2011b,Matsunaga2013}
will provide the opportunity to use the same stellar tracer to constrain the change
in iron abundance across the corotation resonance. This also means the opportunity
to constrain whether the high star formation rate of the innermost Galactic regions
is driven by a disk instability that is dragging material from the inner disk into
these regions \citep{Freeman2013,Ness2013a,Ness2013b}.

Classical Cepheids are also excellent tracers to constrain the speed of the spiral 
arm pattern by fitting a kinematic model to the observed Cepheid kinematics 
\citep{Fernandez2001,Lepine2001}. The Cepheid kinematics is time consuming, 
since a proper coverage of the radial 
velocity curves does require spectroscopic time series data. The use of template 
radial velocity curves significantly decreases the number of measurements required 
for an accurate estimate of the center of mass radial velocity \citep{Metzger1998}. 
However, we still lack accurate radial velocity curve templates covering the entire 
period range.  

Current observational scenario appears to be even more appealing in the outer disk,
since we are still facing a "Cepheid desert" for Galactocentric distances larger 
than $\sim$18 kpc. New identification and characterization of Cepheids at least 
in the first and in the second quadrant are urgently needed to properly trace 
the outskirts of the Galactic disk.

\begin{acknowledgements}
It is a pleasure to thank M. Zoccali and N. Suntzeff for many interesting 
discussions concerning the bulge and the thin disk metallicity gradients.  
We also acknowledge an anonymous referee for his/her pertinent suggestions 
that improved the content and the readability of the paper.  
We are very grateful to VLT staff astronomers for transforming the original 
observing proposals into a solid experiment.  
This work was partially supported by PRIN--INAF 2011 "Tracing the
formation and evolution of the Galactic halo with VST" (P.I.: M. Marconi)
and by PRIN--MIUR (2010LY5N2T) "Chemical and dynamical evolution of
the Milky Way and Local Group galaxies" (P.I.: F. Matteucci).
One of us (K.G.) thank the ESO for support as science visitor, G.B. thanks 
The Carnegie Observatories visitor programme for support as science visitor.
This research made use of spectra obtained from the ESO Science Archive Facility. 
This publication makes use of data products from the Two Micron All Sky Survey, 
which is a joint project of the University of Massachusetts and the Infrared 
Processing and Analysis Center/California Institute of Technology, funded by 
the National Aeronautics and Space Administration and the National Science 
Foundation. This research has made use of the WEBDA database, operated at 
the Institute for Astronomy of the University of Vienna.
\end{acknowledgements}



%
%
%
 \begin{figure*}
 \centering
 \includegraphics[width=7cm]{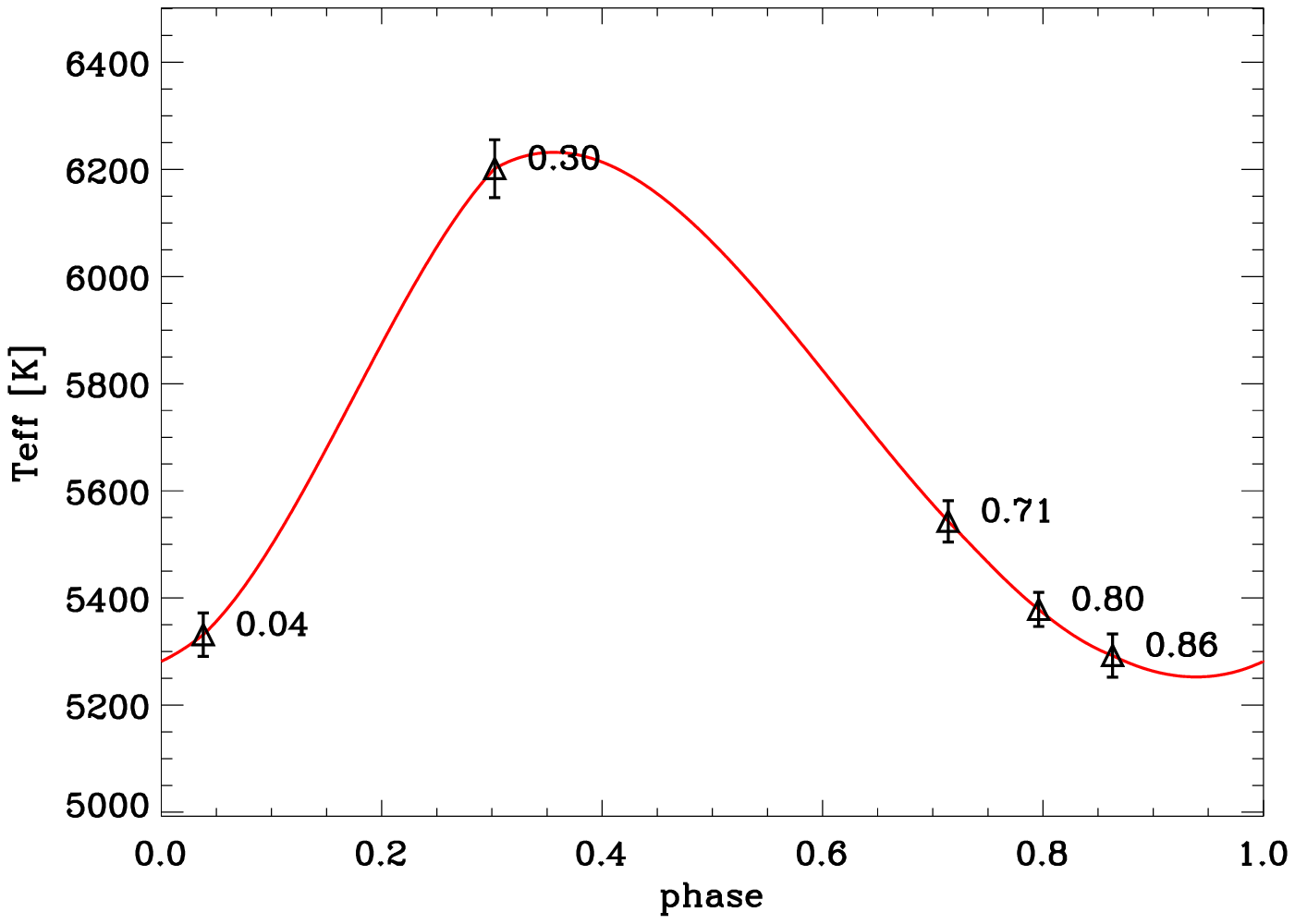}
 \includegraphics[width=7cm]{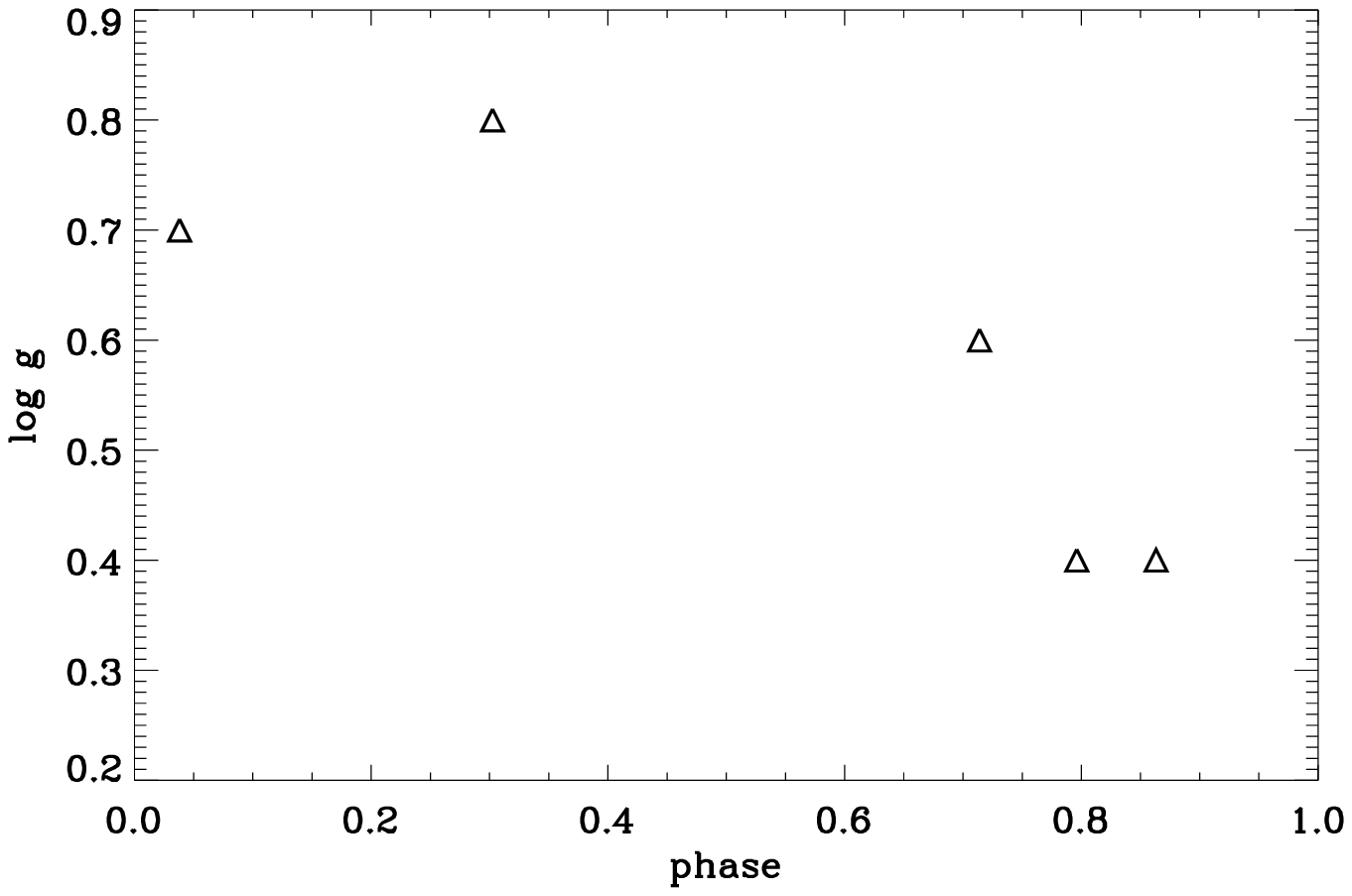}
 \caption{Top -- Effective temperature curve versus pulsational phase for 
 the Cepheid RY Sco as derived using the LDRs method (see text for details). 
 The adopted spectra are the same of Fig.~\ref{fig:fig1}. The \teff \ 
 estimates derived for each spectrum and its error, are plotted 
 as black triangles. The red line shows the spline fit. The number plotted 
 on top of the individual measurements show the pulsation phase. 
 Bottom -- Same as the top, for the surface gravity. \label{fig:RYSco}}
 \end{figure*}
%
%
 \begin{figure*}
 \centering
 \includegraphics[width=15cm]{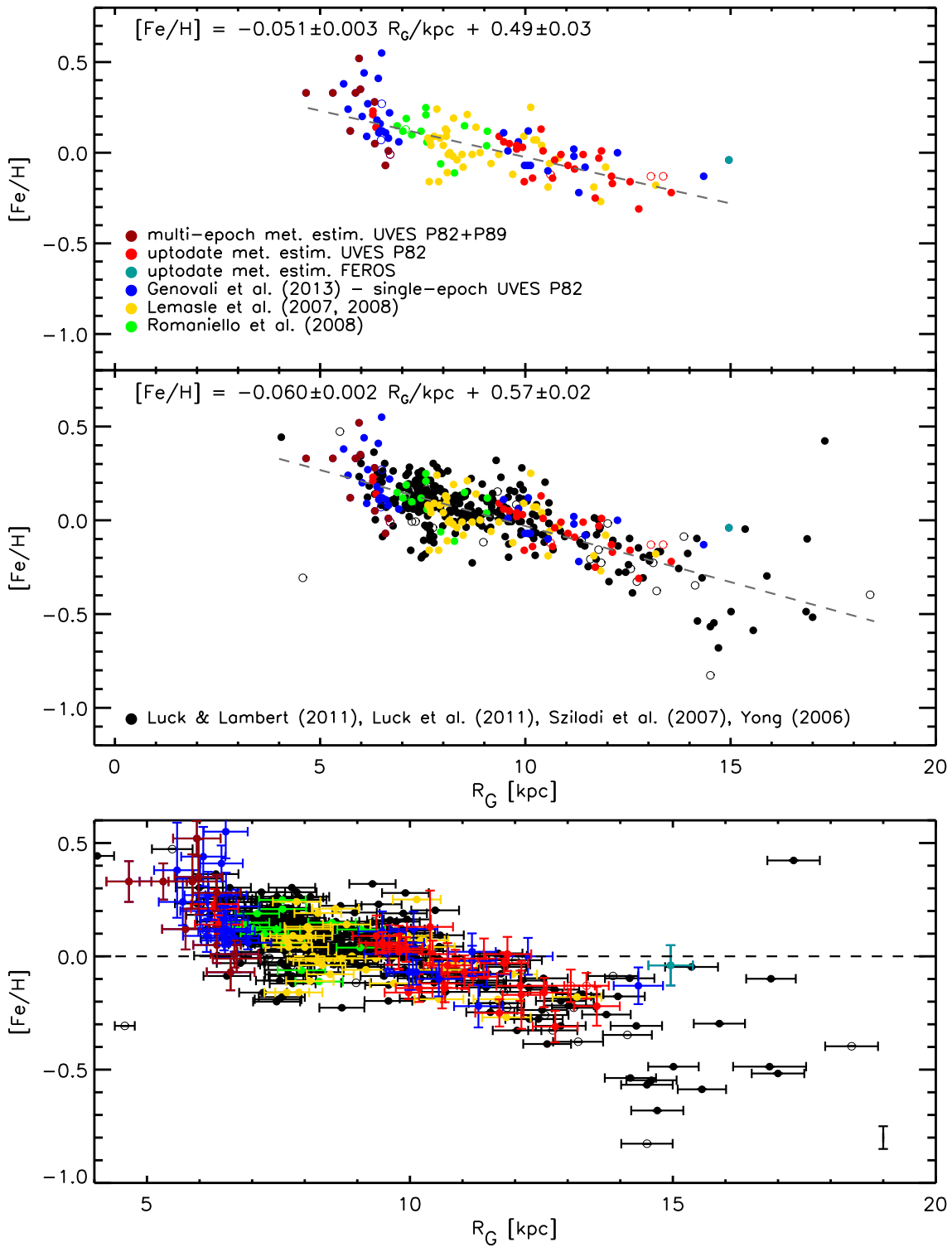}
 \caption{
 Top -- Iron abundances of Galactic classical Cepheids versus 
 Galactocentric distance. Spectroscopic measurements based on different 
 data sets are plotted with different colors. They include 
 updated iron abundances based on current UVES spectra (30, red) and 
 on multi-epoch UVES spectra (11, dark red); plus updated iron abundance 
 for CE Pup based on a FEROS spectrum (light blue)  and iron abundances 
 provided by our group: \cite{Genovali2013} (33, blue), \cite{Lemasle2007}, 
 \cite{Lemasle2008} (39, yellow), \cite{Romaniello2008} (14, green). 
 Cepheids that according to the General Catalog of Variable Stars 
 \citep{Samus2009} are candidate classical Cepheids were plotted with 
 open circles. The grey dashed line shows the metallicity gradient. 
 Middle -- Same as the top, but the iron abundances include our 
 measurements and those available in the literature: \cite{Luck2011a},
 \cite{Luck2011b}, and \cite{Sziladi2007} (322, black).  
 The grey dashed line shows the metallicity gradient based on the 
entire sample. 
 Bottom --  Zoom of the top panel for Galactocentric distances ranging 
 from 4 to 20 kpc. The bars on individual Cepheids display 
 the uncertainty both on iron abundance and on distance. The vertical black 
 bar on the bottom right corner shows the mean uncertainty on 
 \cite{Luck2011b} and \cite{Romaniello2008} iron abundances.
\label{fig:grad}}
 \end{figure*}
%
 \begin{figure*}
 \centering
 \includegraphics[width=15cm, height=10truecm]{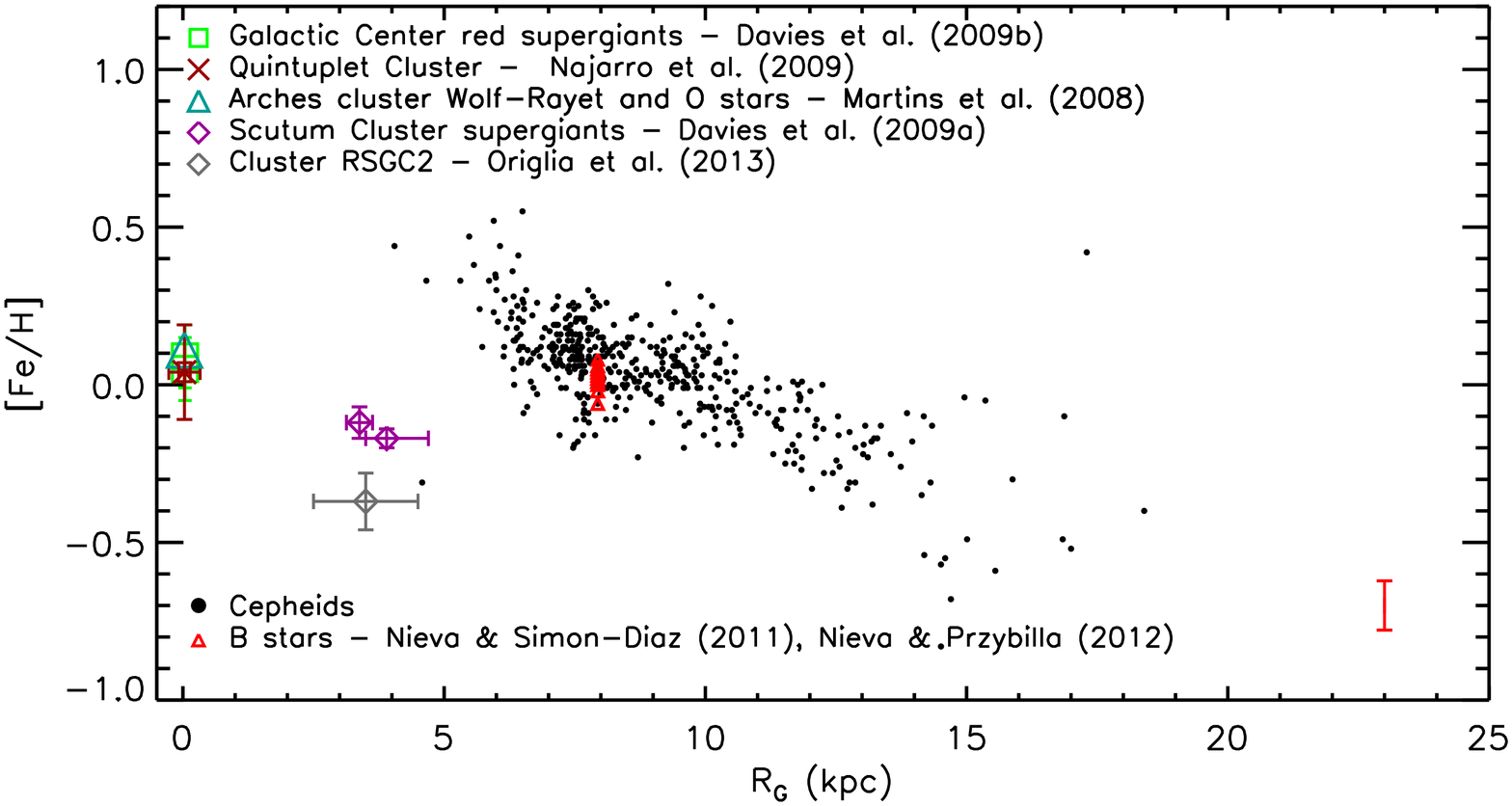}
 \caption{
 Comparison of the Cepheid (black dots) metallicity 
 gradient with iron abundances of early B--type stars 
 located either in the solar neighborhood or in the Orion star 
 forming region \citep{Nieva2011,Nieva2012}. 
 The green square marks the iron abundance of the two red supergiants 
 in the Galactic center measured by \citep{Davies2009a}, while the magenta 
 diamonds the 26 red supergiants in the Scutum Clusters measured by 
 \citep{Davies2009b}, the grey diamond the three red supergiants 
 in the cluster RSGC2 measured by \citep{Origlia2013}, the 
 red cross the two luminous blue variables 
 (LBVs) in the Quintuplet cluster \citep{Najarro2009} and the 
 light-blue triangle three Wolf--Rayet and two O-type stars in the 
 Arches cluster \citep{Martins2008}. The red vertical bar on the bottom 
 right corner shows the mean uncertainty on iron abundances. 
 \label{fig:plotHII_OBs}}
 \end{figure*}
%
%
 \begin{figure*}
 \centering
 \includegraphics[width=12cm]{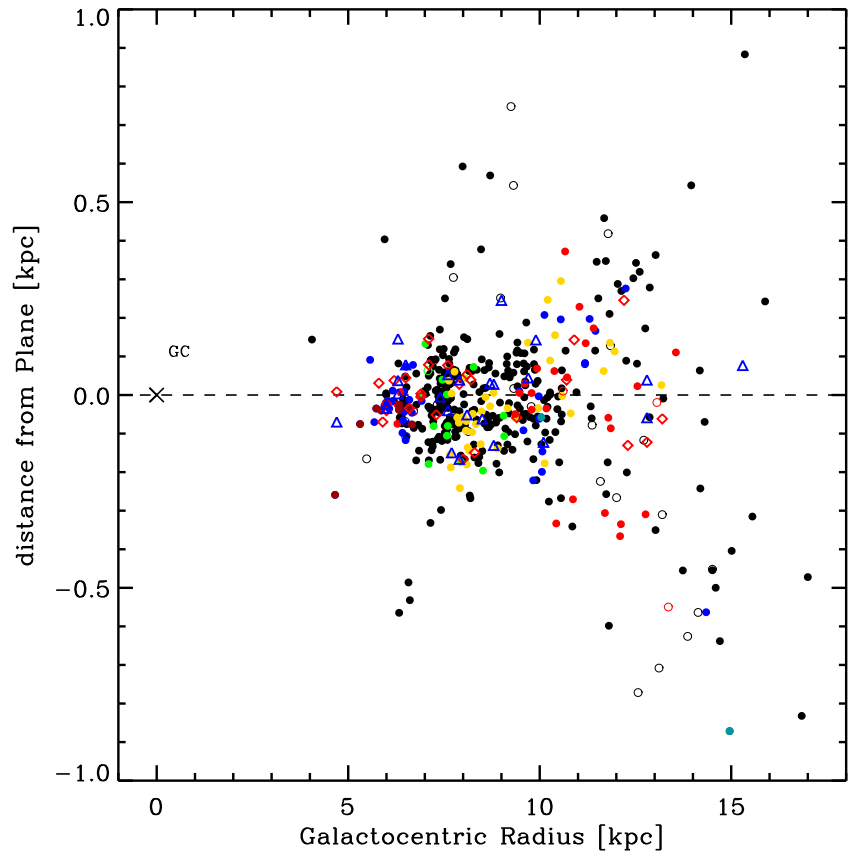}
 \caption{Distance from the Galactic plane versus Galactocentric distance 
 of Galactic Cepheids with accurate iron abundances. 
 Symbols and colors are the same as in Fig.~\ref{fig:grad} the black cross 
 marks the position of the Galactic Center (GC).
 }
 \label{fig:cepazim}
 \end{figure*}

%
%
 \begin{figure*}
 \centering
 \includegraphics[width=15cm, height=15truecm]{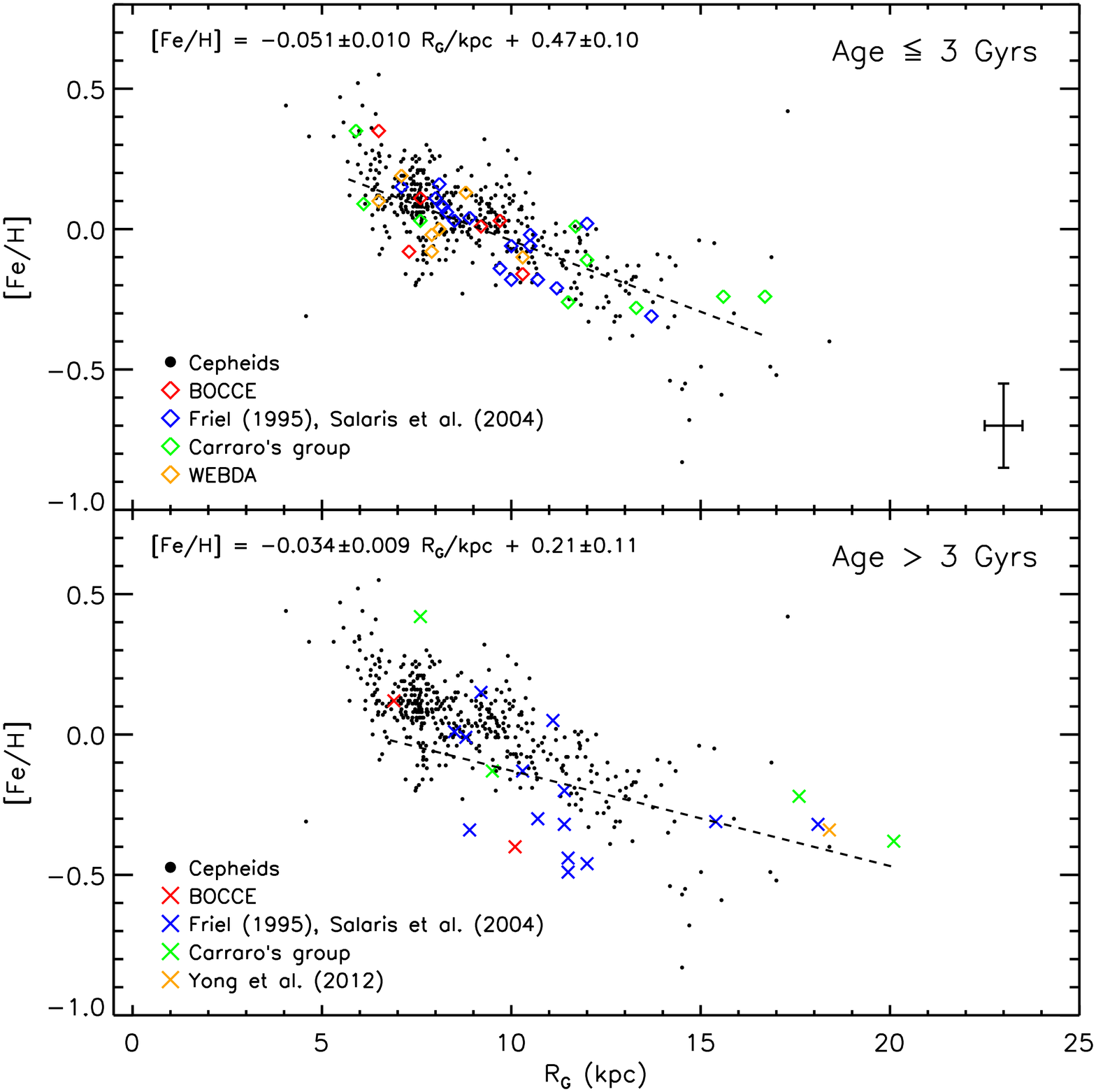}
 \caption{
 Top -- Comparison of the Cepheid (black dots) metallicity gradient 
 with iron abundances for open clusters younger than 3 Gyrs. The color 
 coding does refer to different subsamples: 
 BOCCE project (red symbols); cluster ages from \cite{Salaris2004} and 
 distances from \cite{Friel1995} (blue); Carraro et al.  (green symbols); 
 WEBDA (yellow symbols).  See Table~\ref{tab:OCs} for individual values and 
 references. 
 The vertical and horizontal black error bars on the bottom right corner 
 show the mean uncertainty on Galactocentric distances and iron abundances 
 for OCs in the above subsamples.
 The dashed line shows the metallicity gradient based on the selected 
 open clusters.  
 Bottom -- Same as the top, but for cluster ages older than 3 Gyrs. 
 The yellow cross marks the position of the subsample from \cite{Yong2012}. 
 The dashed line shows the metallicity gradient based on the selected 
 open clusters. 
 \label{fig:plot_OCs}} 
 \end{figure*}
%
 \clearpage
 \begin{figure*}
 \centering
 \includegraphics[width=9cm]{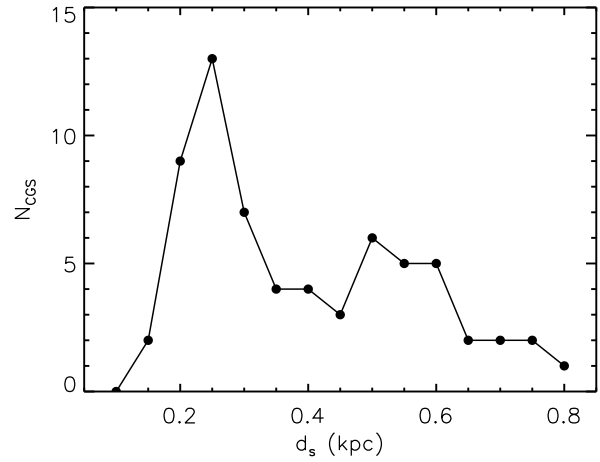}
 \caption{Number of candidate Cepheid groups as a function of the 
 "search distance". See text for more details. \label{fig:fig1_island}}
 \end{figure*}
 
 \begin{figure*}
 \centering
 \includegraphics[width=13cm,angle=180]{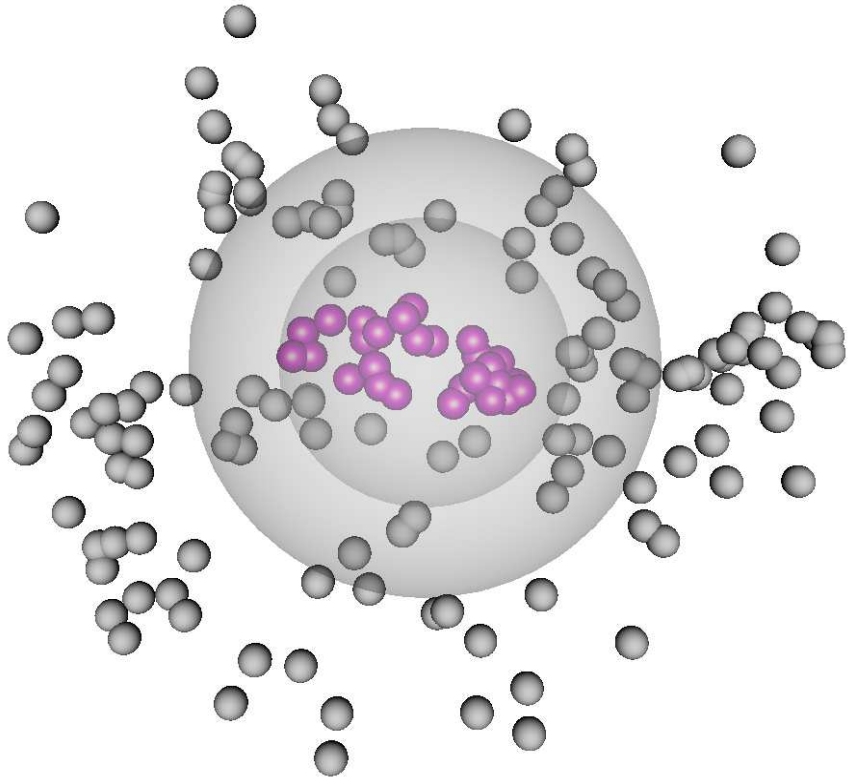}
 \caption{3D plot showing the approach we adopt to select candidate CGs. 
 The members of a generic CG are displayed with magenta sphere, while the 
 surrounding Cepheids with grey spheres. The dark grey shaded inner area  
 shows the sphere including the members of the CG (the radius $R_{CG}$). 
 The light grey shaded outer area shows the spherical layer with a 
 volume three times larger than the volume of the selected CG. 
 Candidate CGs have a Cepheid density that is 3.5 larger than the average 
 stellar density of its neighborhood. \label{fig:sfere3g}}
 \end{figure*}
%
 \begin{figure*}
 \centering
 \includegraphics[width=12cm]{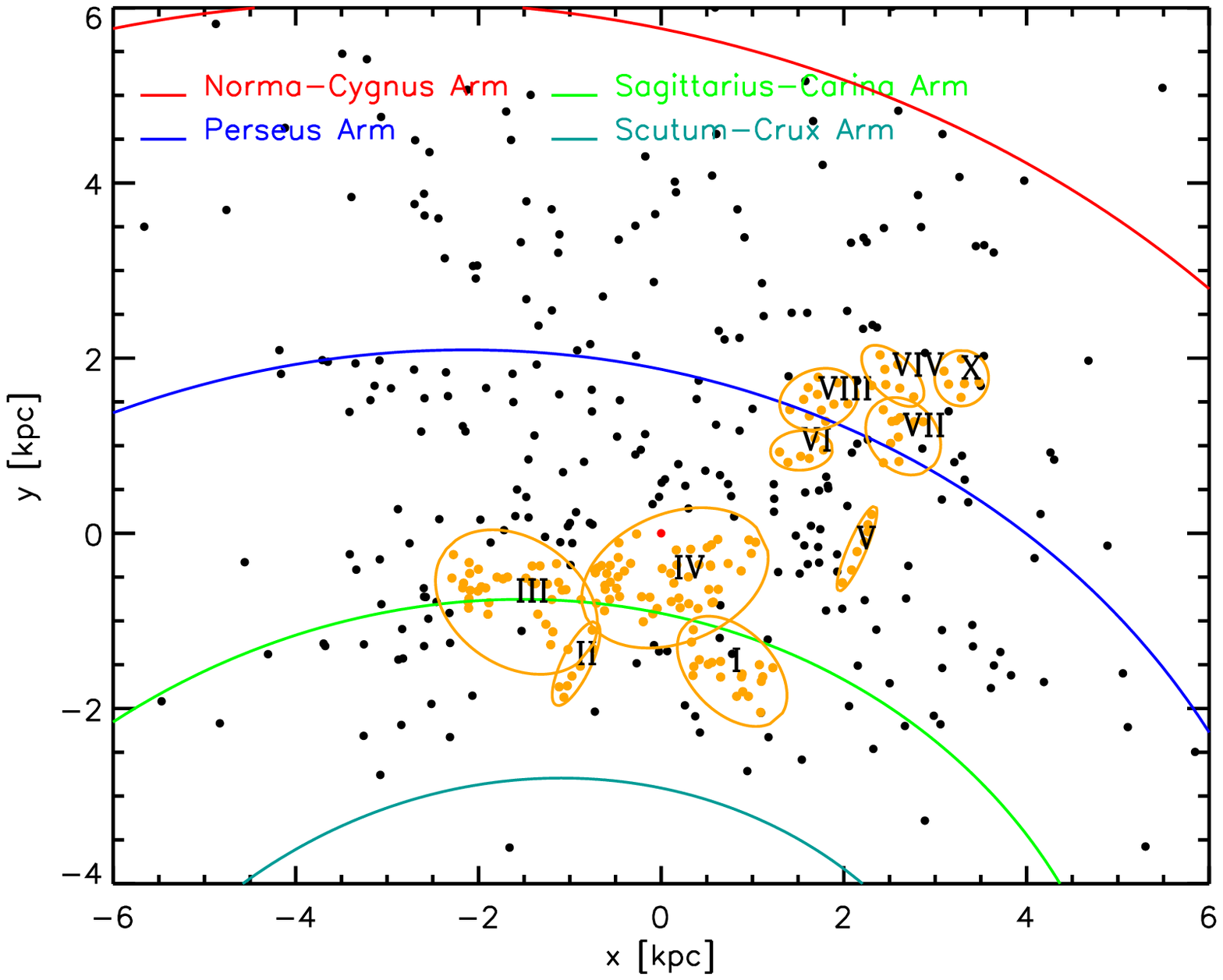}
 \includegraphics[width=12cm]{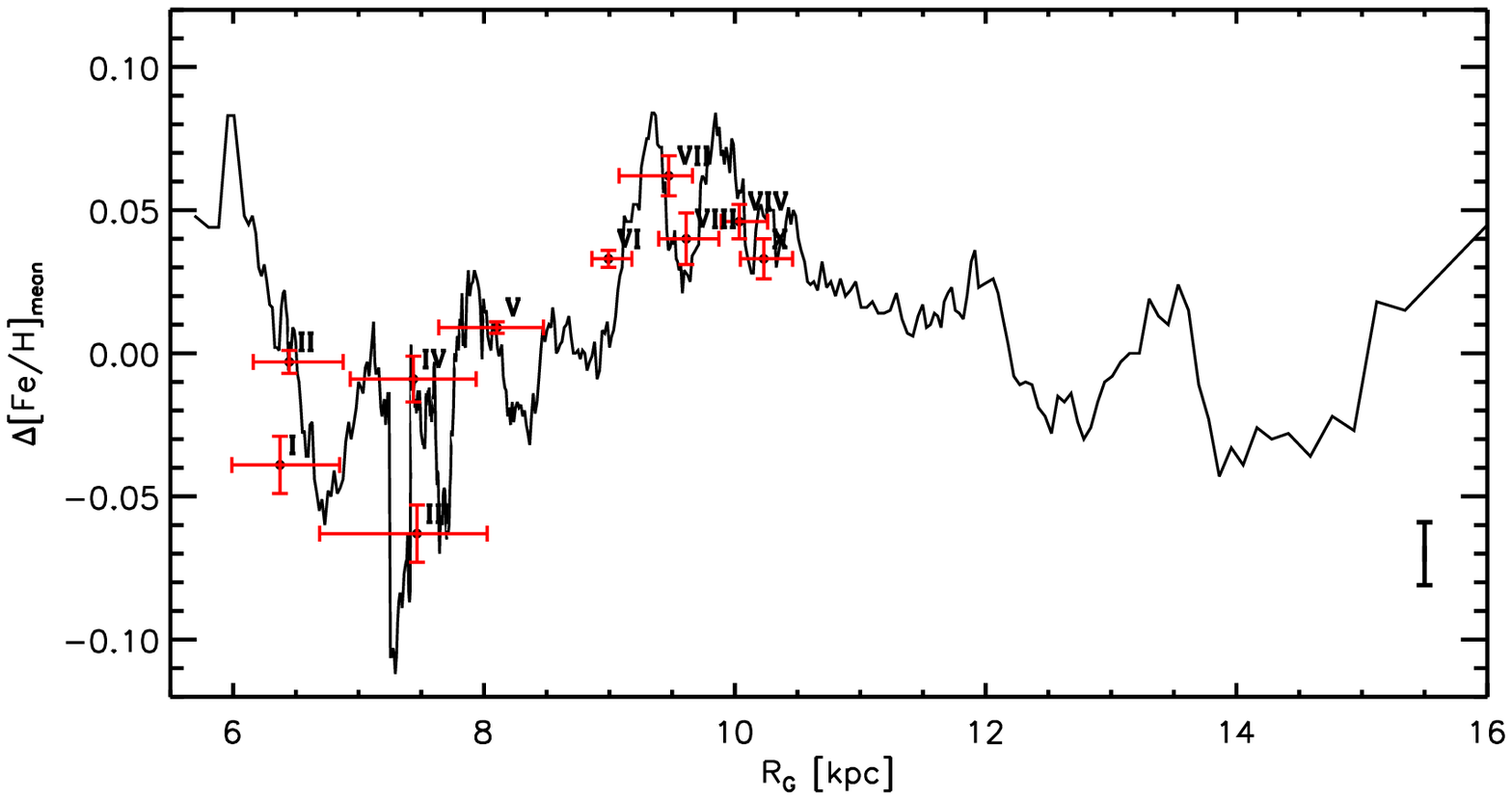}
 \caption{Top -- Projection onto the Galactic plane of isolated Cepheids 
 (black dots) and Cepheids members of candidate CGs (yellow dots). The ellipses mark 
 the edges of individual CGs. They are also labeled with the identification 
 number given in Table~\ref{tab:islands}. The red dot shows the Sun position. 
 The colored lines display a logarithmic model of the spiral arms 
 \cite[][see text for details]{Vallee2002,Cordes2002}. The names of the spiral 
 arms are labeled.  
 Bottom --  Running average of Cepheids metallicity residuals versus Galactocentric 
 radius (black line). The red dots mark the mean metallicity of the candidate 
 CGs once the metallicity gradient has been subtracted. The red vertical lines 
 display the intrinsic dispersion in iron of candidate CGs. The red horizontal lines show the 
 inner and the outer Galactocentric distance of individual candidate CGs 
 (see columns 6 and 7 in Table~\ref{tab:islands}).
 \label{fig:fig23_island}}
 \end{figure*}
%
%
%
%
 \begin{figure*}
 \centering
 \includegraphics[width=8cm]{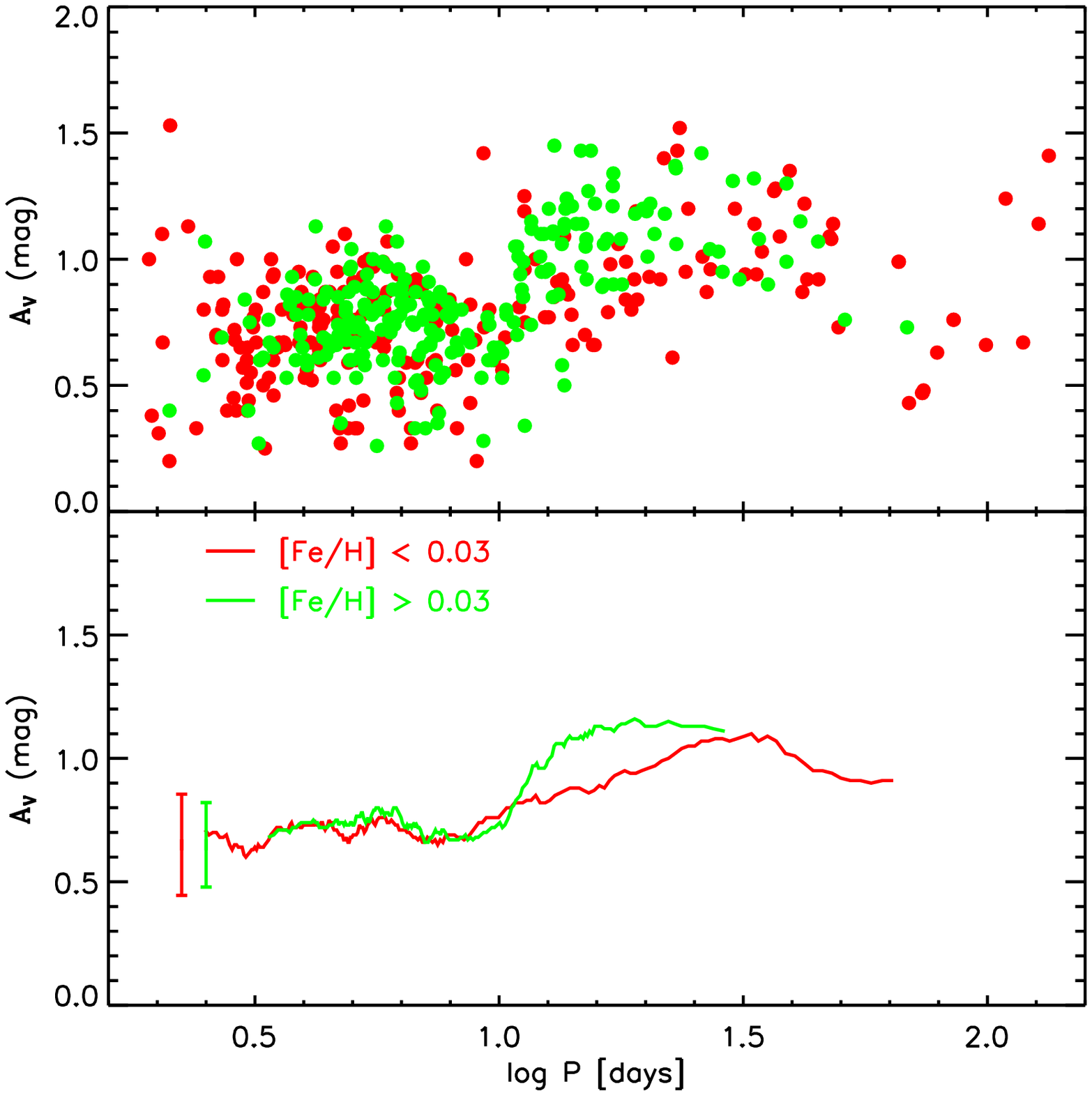}\hspace*{0.5truecm}\includegraphics[width=8cm]{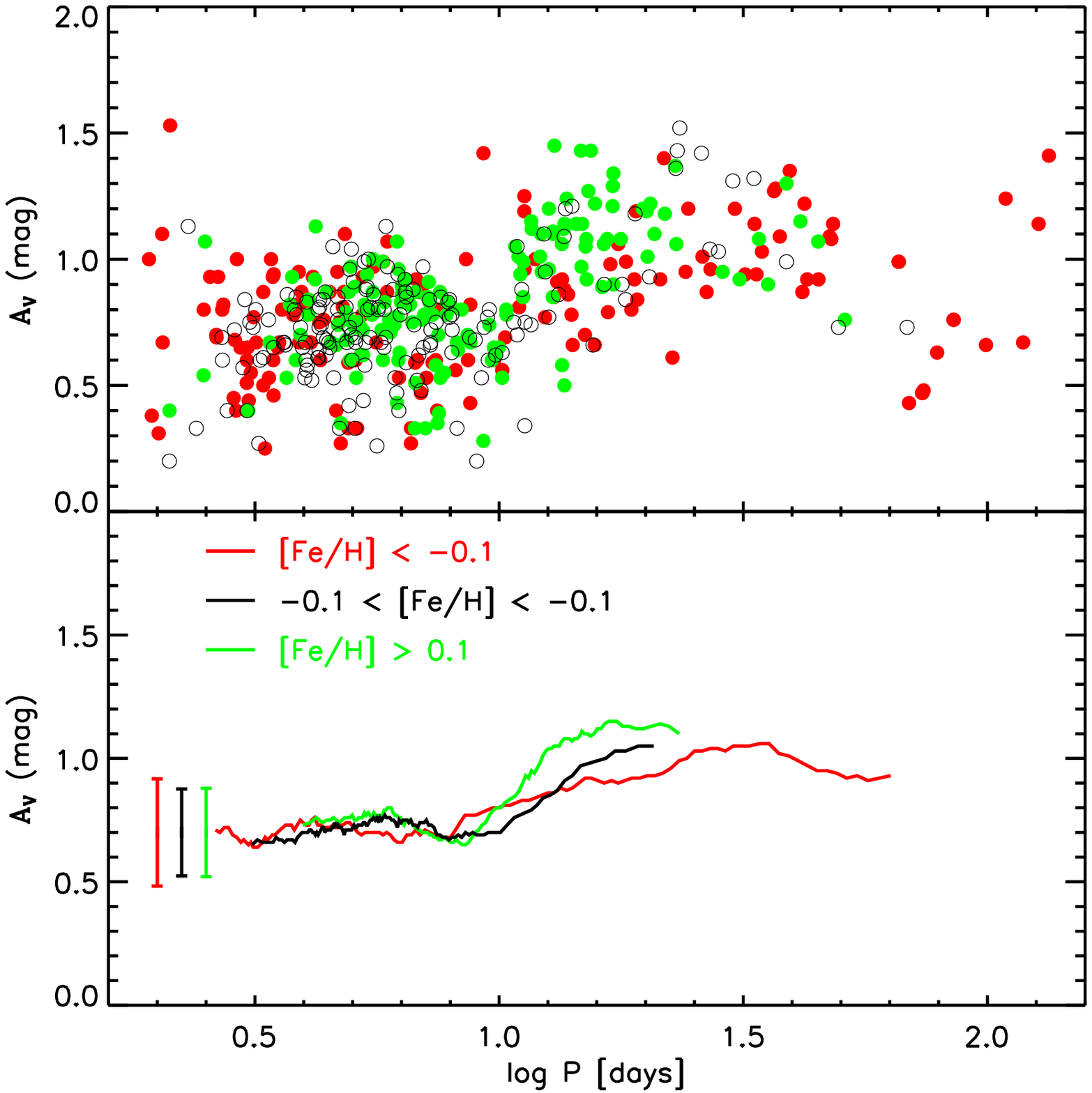}
 \caption{Visual amplitudes versus logarithmic period for Galactic and 
 Magellanic Cepheids. Different colors indicate different metallicity 
 ranges. 
 Top left -- the Cepheid sample was divided into two sub-samples: 
 super-solar (green circles) and sub-solar (red) metallicity. 
 Bottom left -- Running mean of the luminosity amplitudes plotted 
 in the top panel. The color coding is the same of the top panel. 
 The vertical bars display the intrinsic error of the running mean 
 due to bin size and number of stepping stars (see text for more 
 details). 
 Top right -- Same as the top left panel, but the Cepheid sample was 
 divided into three subsamples: super-solar (green), 
 solar (empty black circles), and sub-solar (red) metallicity. 
 Bottom right -- Same as the bottom left panel, but for the three 
 different metallicity bins adopted in the top panel. \label{fig:amp}}
 \end{figure*}
%
%

\onecolumn 
\clearpage
\begin{center} 
\scriptsize 
\begin{longtable}{rccrrccccrcrl}
\caption{Intrinsic parameters and abundances for the current sample of Classical Cepheids based on individual spectra.}\label{tab:table_atm_par}\\ 
\hline
name  & date  & MJD & ET & \snr & \teff & \logg & \vt     & \fei & N & \feii & N &  Spect.  \\
      &       &     & (s)&      &   $K$ &       &  (\kms) &      &   &       &   &               \\
\hline \hline
 V340  Ara   & 2012-07-29   &    56137.137   &    120   &    99   &  5210 $\pm$   49   &  0.30   & 5.0   &  0.28 $\pm$ 0.27   &  23   &  0.27 $\pm$ 0.11    &   2   &   UVES   \\
 V340  Ara   & 2008-08-30   &    54708.065   &    300   &   241   &  5210 $\pm$   88   &  0.00   & 3.0   &  0.53 $\pm$ 0.15   &  53   &  0.53 $\pm$ 0.11    &   4   &   UVES   \\
 V340  Ara   & 2008-08-31   &    54709.079   &    300   &   264   &  5130 $\pm$   96   &  0.40   & 3.0   &  0.42 $\pm$ 0.22   &  26   &  0.65 $\pm$ 0.24    &   3   &   UVES   \\
 V340  Ara   & 2012-07-30   &    56138.094   &    120   &   122   &  5610 $\pm$   68   &  0.70   & 4.4   &  0.30 $\pm$ 0.20   &  41   &  0.33 $\pm$ 0.10    &   2   &   UVES   \\
 V340  Ara   & 2012-07-31   &    56139.185   &    120   &   114   &  6060 $\pm$   65   &  0.80   & 5.2   &  0.24 $\pm$ 0.13   &  51   &  0.22 $\pm$ 0.01    &   2   &   UVES   \\
\hline \hline
\end{longtable}
\tablefoot{From left to right the columns give the Cepheid name, the date of acquisition and 
the modified Julian Day of the spectrum, the exposure time, maximum \snr \ of the spectrum, 
the estimated intrinsic parameters (\teff, \logg, \vt). Columns 9 and 10 list the \fei \ 
abundance and its standard deviation together with the number of \fei \ lines measured. 
Columns 11 and 12 list the  \feii \ abundance and its standard deviation together with 
the number of \feii \ lines measured. The last column gives the name of the adopted spectrograph.}
\end{center}

\clearpage 
\vspace*{0.5truecm} 
\begin{table}[htp!]
\scriptsize 
\caption{Impact of uncertainties affecting intrinsic parameters on iron abundance.}
\label{tab:error}
\centering
\begin{tabular}{lrrrr l|r|r|r|r|r|r|}
\hline
name &	\teff & \feh \ & \logg & \vt & element & \multicolumn{2}{c |}{$\Delta$\teff } & \multicolumn{2}{c|}{$\Delta$\logg } & \multicolumn{2}{ c |}{$\Delta$\vt} \\
     &       &        &      &    &         &     -100 K     & +100 K     &     -0.3   &    +0.3      &       -0.5    &   +0.5 \\
\hline
\hline
V340 Ara  &	6060 &	 0.22 $\pm$ 0.01 &	0.8 &	5.2 &		   \fei \ &	    0.17 &	 0.31 &	 0.28 & 	0.22 & 	0.27 &	 0.23 \\		
			&		 &     &    &   &                         \feii \ &	    0.25 &	 0.20 &	 0.12 &	 0.33 & 	0.26 &	 0.20 \\		
				&		 &     &    &   &                    \feh \ &	    0.21 &	 0.26 &	 0.20 &	 0.28 & 	0.27 &	 0.22 \\
\hline
V340 Ara &	4820 &	 0.23 $\pm$	0.18 &	1.0 &	4.4 &		   \fei \ &	    0.14 &	 0.30 &  0.20 &	 0.21 & 	0.27 &	 0.15	\\
			&		 &     &    &   &                         \feii \ &	    0.38 &	 0.22 &	 0.14 &	 0.46 & 	0.42 &	 0.20 \\		
				&		 &     &    &   &                    \feh \ &	    0.26 &	 0.26 &	 0.17 &	 0.34 & 	0.35 &	 0.18 \\
\hline
V500 Sco &	5280 &	-0.03 $\pm$	0.12 &	0.4 &	3.8 &		   \fei \ &	   -0.11 &	 0.06 &	 0.01 &	-0.07 & 	0.04 &	-0.07 \\		
			&		 &     &    &   &                         \feii \ &	    0.00 & -0.07 &	-0.16 &	 0.09 &	 0.02 &	-0.08	\\	
				&		 &     &    &   &                    \feh \ &	   -0.06 &	-0.01 &	-0.08 &	 0.01 &	 0.03 &	-0.08 \\
\hline
V500 Sco &	6020 &	-0.11 $\pm$	0.07 &	1.1 &	4.0 &		   \fei \ &	   -0.18 &	-0.05 &	-0.09 &	-0.13 &	-0.08 &	-0.13	\\	
			&		 &     &    &   &                         \feii \ &	   -0.11 &	-0.11 &	-0.21 &	-0.01 &	-0.05 &	-0.15 \\		
				&		 &     &    &   &                    \feh \ &	   -0.15 &	-0.08 &	-0.15 &	-0.07 &	-0.07 &	-0.14 \\
\hline
V510 Mon &	5505 &	-0.16 $\pm$	0.06 &	0.9 &	3.4 &	   	\fei \ &	   -0.25 &	-0.07 &	-0.16 &	-0.08 &	-0.08 &	-0.19 \\		
			&		 &     &    &   &                         \feii \ &	   -0.14 &	-0.19 &	-0.05 &	-0.10 & -0.10 & -0.21	\\	
				&		 &     &    &   &                    \feh \ &	   -0.20 &	-0.13 &	-0.11 &	-0.09 &	-0.09 &	-0.20 \\
\hline
\hline
\end{tabular}
\tablefoot{Error budget for V340 Ara, V500 Sco and V510 Mon. Columns 7 to 12 list plausible changes in 
atmospheric parameters: effective temperature (\teff), surface gravity (\logg) and microturbulent velocity (\vt) 
together with their impact on \fei \ and \feii \ abundances. The mean \feh \ abundance for each set of intrinsic 
parameters is also given.}
\end{table}

\clearpage
\begin{center} 
\scriptsize 
\begin{longtable}{r lc ccc cc cc ccl}
\caption{Mean near-infrared magnitudes, mean distances and mean iron abundances for the current sample of Classical Cepheids.}\label{tab:tab_distances}\\
\hline
Name  & Type & log($P$)  & $<J>$& $<H>$& $<K>$& $[Fe/H]_{lit}$  & $[Fe/H]$  & $N_S$& $\mu^d$  & $\Rg ^e$ & Notes & Ref.$^f$ \\
      &      & [days]    & mag  & mag  & mag  &                 &           &            & mag      & (pc)      &       &     \\
\hline \hline
 V340  Ara &     DCEP &       1.3183   &  7.302   &  6.754   &  6.534   &  0.40   &  0.33$\pm$ 0.09 &  6 & 12.99$\pm$ 0.05   &  4657$\pm$  427     & b$^{\dagger}$   &        PED   \\
   AO  Cma &     DCEP &       0.7646   &  9.154   &  8.595   &  8.366   & -0.04   &  0.01$\pm$ 0.06 &  1 & 12.95$\pm$ 0.05   & 10430$\pm$  433     & b$^{\dagger}$   &        LEM   \\
   TW  Cma &     DCEP &       0.8448   &  7.567   &  7.178   &  7.008   & -0.51   &  0.04$\pm$ 0.09 &  1 & 12.02$\pm$ 0.05   &  9788$\pm$  445     & a               &        LEM   \\
   AD  Gem &     DCEP &       0.5784   &  8.448   &  8.151   &  8.027   & -0.19   & -0.14$\pm$ 0.06 &  1 & 12.24$\pm$ 0.05   & 10662$\pm$  455     & a               &        LEM   \\
   BB  Gem &     DCEP &       0.3633   &  9.696   &  9.335   &  9.215   & -0.10   & -0.09$\pm$ 0.04 &  1 & 12.66$\pm$ 0.07   & 11199$\pm$  460     & c               &       LIII   \\
   DX  Gem &    DCEPS &       0.4966   &  8.759   &  8.336   &  8.207   & -0.04   & -0.01$\pm$ 0.09 &  1 & 12.78$\pm$ 0.09   & 11407$\pm$  473     & c*              &       LIII   \\
    RZ Gem &    DCEP' &       0.7427   &  7.600   &  7.165   &  6.950   & -0.44   & -0.16$\pm$ 0.03 &  1 & 11.56$\pm$ 0.05   &  9973$\pm$  454     & a               &        LEM   \\ 
   BE  Mon &     DCEP &       0.4322   &  8.256   &  7.854   &  7.676   & -0.07   &  0.05$\pm$ 0.09 &  1 & 11.28$\pm$ 0.05   &  9609$\pm$  452                 & a   &        LEM   \\
   CV  Mon &     DCEP &       0.7307   &  7.314   &  6.781   &  6.529   & -0.10   &  0.09$\pm$ 0.09 &  1 & 11.00$\pm$ 0.05   &  9362$\pm$  452                 & a   &        LEM   \\
   TW  Mon &      CEP &       0.8511   &  9.709   &  9.137   &  8.904   & -0.15   & -0.13$\pm$ 0.07 &  1 & 13.76$\pm$ 0.05   & 13059$\pm$  457     & b$^{\dagger}$   &        LEM   \\
   TX  Mon &     DCEP &       0.9396   &  8.569   &  8.114   &  7.922   & -0.12   & -0.03$\pm$ 0.05 &  1 & 13.19$\pm$ 0.05   & 11790$\pm$  452     & a$^{\dagger}$   &        LEM   \\
 V465  Mon &     DCEP &       0.4335   &  8.751   &  8.440   &  8.315   &  0.02   & -0.07$\pm$ 0.07 &  1 & 12.75$\pm$ 0.05   & 11037$\pm$  450                 & a   &       LIII   \\
 V495  Mon &     DCEP &       0.6124   &  9.827   &  9.351   &  9.185   & -0.17   & -0.13$\pm$ 0.07 &  1 & 13.35$\pm$ 0.05   & 12098$\pm$  453     & a$^{\dagger}$   &        LEM   \\
 V508  Mon &     DCEP &       0.6163   &  8.627   &  8.277   &  8.136   & -0.25   & -0.04$\pm$ 0.10 &  1 & 12.42$\pm$ 0.05   & 10714$\pm$  452                 & a   &        LEM   \\
 V510  Mon &     DCEP &       0.8637   &  9.507   &  8.898   &  8.648   & -0.12   & -0.16$\pm$ 0.06 &  1 & 13.51$\pm$ 0.05   & 12550$\pm$  456     & b$^{\dagger}$   &        LEM   \\
   XX  Mon &     DCEP &       0.7369   &  9.402   &  8.903   &  8.699   & -0.18   &  0.01$\pm$ 0.08 &  1 & 13.25$\pm$ 0.05   & 11854$\pm$  451     & b$^{\dagger}$   &        LEM   \\
   QZ  Nor &    DCEPS &       0.5782   &  7.085   &  6.748   &  6.614   &  0.19   &  0.21$\pm$ 0.06 &  2 & 11.53$\pm$ 0.05   &  6283$\pm$  447                 & b   &        G13   \\
   SY  Nor &     DCEP &       1.1019   &  6.638   &  6.091   &  5.864   &  0.34   &  0.23$\pm$ 0.07 &  2 & 11.59$\pm$ 0.05   &  6286$\pm$  446     & b$^{\dagger}$   &       LIII   \\
   CS  Ori &     DCEP &       0.5899   &  9.331   &  8.954   &  8.790   & -0.19   & -0.25$\pm$ 0.06 &  1 & 12.96$\pm$ 0.05   & 11701$\pm$  458                 & a   &        LEM   \\
   AQ  Pup &     DCEP &       1.4786   &  6.033   &  5.495   &  5.283   & -0.26   &  0.06$\pm$ 0.05 &  1 & 12.29$\pm$ 0.05   &  9472$\pm$  436                 & b   &        LEM   \\
   BC  Pup &     DCEP &       0.5495   & 11.015   & 10.475   & 10.247   & -0.23   & -0.31$\pm$ 0.07 &  1 & 14.12$\pm$ 0.05   & 12763$\pm$  426     & b$^{\dagger}$   &        LII   \\
   BN  Pup &     DCEP &       1.1359   &  7.563   &  7.090   &  6.910   & -0.03   &  0.03$\pm$ 0.05 &  1 & 12.83$\pm$ 0.05   &  9930$\pm$  428                 & b   &        LEM   \\
   CE  Pup &     DCEP &       1.6949   &  8.402   &  7.816   &  7.581   & -0.04   & -0.04$\pm$ 0.09 &  1 & 15.27$\pm$ 0.07   & 14958$\pm$  422                 & c   &        LII   \\
   CK  Pup &      CEP &       0.8703   & 10.273   &  9.703   &  9.456   & -0.12   & -0.13$\pm$ 0.06 &  2 & 14.37$\pm$ 0.05   & 13357$\pm$  423     & b$^{\dagger}$   &        G13   \\
   HW  Pup &     DCEP &       1.1289   &  9.468   &  8.914   &  8.716   & -0.28   & -0.22$\pm$ 0.09 &  1 & 14.55$\pm$ 0.07   & 13554$\pm$  436                 & c   &        LEM   \\
   LS  Pup &      CEP &       1.1506   &  8.030   &  7.531   &  7.341   & -0.15   & -0.12$\pm$ 0.11 &  1 & 13.29$\pm$ 0.05   & 10610$\pm$  423                 & b   &        ROM   \\
   VW  Pup &     DCEP &       0.6320   &  9.008   &  8.514   &  8.360   & -0.19   & -0.14$\pm$ 0.06 &  1 & 12.58$\pm$ 0.07   & 10175$\pm$  443                 & c   &        LII   \\
   VZ  Pup &     DCEP &       1.3649   &  7.309   &  6.842   &  6.657   & -0.37   & -0.01$\pm$ 0.04 &  1 & 13.35$\pm$ 0.05   & 10867$\pm$  425                 & b   &        LEM   \\
   WW  Pup &     DCEP &       0.7417   &  8.653   &  8.274   &  8.131   & -0.18   &  0.13$\pm$ 0.16 &  1 & 12.82$\pm$ 0.05   & 10382$\pm$  436     & b$^{\dagger}$   &        LII   \\
    X  Pup &     DCEP &       1.4143   &  6.117   &  5.614   &  5.418   &  0.05   &  0.02$\pm$ 0.08 &  1 & 12.24$\pm$ 0.05   &  9788$\pm$  441                 & b   &        ROM   \\
    KQ Sco &     DCEP &       1.4577   &  5.945   &  5.229   &  4.924   &  0.22   &  0.52$\pm$ 0.08 &  5 & 11.67$\pm$ 0.05   &  5948$\pm$  451     & b   &        ROM        \\   
   RY  Sco &     DCEP &       1.3078   &  4.930   &  4.379   &  4.134   &  0.09   &  0.01$\pm$ 0.06 &  5 & 10.54$\pm$ 0.05   &  6663$\pm$  453                 & b   &        LII   \\
 V500  Sco &     DCEP &       0.9693   &  6.041   &  5.532   &  5.330   &  0.01   & -0.07$\pm$ 0.08 &  4 & 10.65$\pm$ 0.05   &  6590$\pm$  453                 & b   &        LII   \\
   RU  Sct &     DCEP &       1.2945   &  5.891   &  5.287   &  5.034   &  0.11   &  0.14$\pm$ 0.04 &  2 & 11.35$\pm$ 0.05   &  6361$\pm$  449                 & a   &       LIII   \\
   UZ  Sct &     DCEP &       1.1686   &  7.405   &  6.749   &  6.461   &  0.35   &  0.33$\pm$ 0.08 &  6 & 12.29$\pm$ 0.05   &  5309$\pm$  448     & a$^{\dagger}$   &        PED   \\
 V367  Sct &   CEP(B) &       0.7989   &  7.605   &  6.955   &  6.651   & -0.01   &  0.05$\pm$ 0.08 &  4 & 11.23$\pm$ 0.05   &  6332$\pm$  451                 & b   &        LII   \\
    Z  Sct &     DCEP &       1.1106   &  6.949   &  6.477   &  6.263   &  0.33   &  0.12$\pm$ 0.09 &  5 & 12.08$\pm$ 0.05   &  5733$\pm$  445                 & a   &       LIII   \\
   AV  Sgr &     DCEP &       1.1879   &  6.878   &  6.089   &  5.730   &  0.27   &  0.35$\pm$ 0.17 &  5 & 11.49$\pm$ 0.05   &  5980$\pm$  454     & b $^{\dagger}$  &       PED    \\
   VY  Sgr &     DCEP &       1.1322   &  7.156   &  6.385   &  6.042   &  0.35   &  0.33$\pm$ 0.12 &  5 & 11.63$\pm$ 0.05   &  5862$\pm$  453     & b$^{\dagger}$   &        PED   \\
   WZ  Sgr &     DCEP &       1.3394   &  5.255   &  4.752   &  4.536   &  0.19   &  0.28$\pm$ 0.08 &  5 & 11.10$\pm$ 0.05   &  6326$\pm$  453     & a               &        LII   \\
   XX  Sgr &      CEP &       0.8078   &  6.436   &  5.967   &  5.747   &  0.10   & -0.01$\pm$ 0.06 &  5 & 10.55$\pm$ 0.07   &  6706$\pm$  453     & c               &      LII   \\
   EZ  Vel &     DCEP &       1.5383   &  8.884   &  8.203   &  7.914   & -0.01   & -0.17$\pm$ 0.15 &  1 & 14.97$\pm$ 0.07   & 12119$\pm$  358     & c               &      LEM   \\
\hline \hline
\end{longtable}
\vspace*{12.15truecm}\hspace*{-5truecm}\tablefoot{
From left to right the columns give the target name, the variable type 
(from GCVS), the logarithmic period, the mean magnitudes in  \textit{J}, \textit{H} and 
\textit{K} bands. Columns 7 and 8 give the iron abundance available in the 
literature re-scaled to our solar abundance, and the current iron abundance. 
Column 9 gives the number of spectra used to estimate the mean iron abundance.
Columns 10 and 11 list the true distance modulus and the 
Galactocentric distance. The last two columns give the notes on the 
NIR photometry and distances for individual objects and the references. 

\textbf{(a)} Mean magnitudes provided by \cite{Monson2011} transformed into 
the 2MASS photometric system using the zero-points given in their Table~1.  
\textbf{(b)} Mean magnitudes provided by \cite{Laney1992} and by Laney 
(private comm.) transformed into the 2MASS photometric system using the 
zero--points provided by \cite{Koen2007}. The objects marked with a 
$^{\dagger}$ (Laney, private comm.) do not have a complete coverage of 
the light-curve (the number of phase points ranges from 4 to 14). 
\textbf{(c)} Mean magnitudes based on single-epoch measurements from the 
2MASS catalogue and the NIR template light curves provided by 
\cite{Soszynski2005}. An asterisk indicates the use of single-epoch 
photometry in the distance determination (see section \ref{sec:distances} 
for details). For these stars the magnitudes listed in columns 4, 5 and 6 
are the single-epoch measurements retrieved from the 2MASS catalog. 
\textbf{(d)} The weighted mean of the three true distance moduli. 
The errors account for uncertainties affecting the mean magnitudes 
and for the intrinsic dispersion of the adopted NIR PW relations.
\textbf{(e)} The weighted mean Galactocentric distances were estimated assuming 
$\Rg$=7.94$\pm$0.37$\pm$0.26 kpc \citep{Groenewegen2008}. The errors account 
for uncertainties affecting both the solar Galactocentric distance and 
the heliocentric distances.
\textbf{(f)} References for the iron estimate given in column 7. 
The priority was given, in the following order, to evalutions 
from our group (G13: \cite{Genovali2013}, PED: \cite{Pedicelli2010}, 
LEM: \cite{Lemasle2007,Lemasle2008}, ROM: \cite{Romaniello2008}) 
and from the literature: LII: \cite{Luck2011a}, LIII: \cite{Luck2011b}.
}
\end{center} 

\clearpage
\begin{center} 
\scriptsize 
\begin{longtable}{r lc ccc cc cc ccl}
\caption{Galactic Cepheids for which the iron abundance was available in the literature.}\label{tab:tab_distances_all}\\
\hline
Name  & Type & log($P$)  & $<J>$  & $<H>$  & $<K>$  & $[Fe/H]_{lit}$  & $[Fe/H]$  & $\mu^d$  & $\Rg ^e$ & Notes & Ref.$^f$ \\
      &      & [days]    & mag  & mag  & mag  &   &   & mag & (pc) &   &     \\
\hline \hline
    T  Ant &     DCEP &       0.7707   &  7.914   &  7.602   &  7.494   & -0.20   & -0.23   & 12.35$\pm$ 0.05   &  8706$\pm$  427                 & b   &       LIII   \\
   BC  Aql &      CEP &       0.4632   & 11.634   & 11.283   & 11.180   & -0.28   & -0.31   & 14.97$\pm$ 0.07   &  4578$\pm$  190                 & c   &       LIII   \\
   EV  Aql &      CEP &       1.5885   &  8.990   &  8.369   &  8.162   &  0.06   &  0.03   & 15.48$\pm$ 0.07   &  9249$\pm$  287                 & c   &       LIII   \\
   FF  Aql &    DCEPS &       0.6504   &  4.030   &  3.741   &  3.525   &  0.04   &  0.10   &  8.71$\pm$ 0.37   &  7592$\pm$  455                & c*   &        LII   \\
   FM  Aql &     DCEP &       0.7863   &  5.657   &  5.210   &  5.007   &  0.24   &  0.21   &  9.76$\pm$ 0.05   &  7326$\pm$  450                 & a   &       LIII   \\
\hline\hline
\end{longtable}
\tablefoot{From left to right the columns give the same quantities of 
Table \ref{tab:tab_distances}. Column 7 gives the original iron estimate available 
in the literature, while column 8 gives the iron abundance re-scaled to current 
metallicity scale. In column 2 is reported the GCVS classification type. 
Note that V556 Cas and V1397 Cyg are misclassified in the GCVS, since they have 
been identified as classical Cepheids by \citep{Wils2004}. Moreover, their 
nature was confirmed by \citep{Luck2011b} on the basis of spectroscopic evidence.
The acronym SZI refers to \cite{Sziladi2007}, while YON to \cite{Yong2006}.

}
\end{center}

\clearpage
\begin{center} 
\scriptsize 
\begin{longtable}{r r r c c r r}
\caption{Individual ages, distances and metallicities for the current sample of open clusters.}\label{tab:OCs}\\
\hline
ID  &  Age    & $\Rg$    & \feh$_{lit}$ & \feh$_{rsc}$ &  Ref. met. & Ref. age-dist. \\
    &  (Gyrs) & (kpc)    &               &             &            &   \\
\hline\hline

          Be 17  &  10.06  &   10.3  &  -0.15  &  -0.13  &            Friel et al. (2005)  &   SF$^a$ \\
          Be 18  &   5.69  &   11.5  &  -0.44  &  -0.44  &             Yong et al. (2012)  &   SF \\
          Be 20  &   4.05  &   15.4  &  -0.30  &  -0.31  &          Sestito et al. (2008)  &   SF \\
          Be 21  &   2.18  &   13.7  &  -0.31  &  -0.31  &             Yong et al. (2012)  &   SF \\
          Be 22  &   4.26  &   11.4  &  -0.45  &  -0.45  &             Yong et al. (2012)  &   SF \\
          Be 25  &   4.00  &   17.6  &  -0.20  &  -0.22  &          Carraro et al. (2007b)  &    C$^b$ \\
          Be 29  &   4.34  &   18.1  &  -0.31  &  -0.32  &          Sestito et al. (2008)  &   SF \\
          Be 31  &   5.32  &   11.5  &  -0.53  &  -0.49  &            Friel et al. (2005)  &   SF \\
          Be 32  &   5.91  &   10.7  &  -0.38  &  -0.38  &             Yong et al. (2012)  &   SF \\
          Be 39  &   7.00  &   11.1  &   0.03  &   0.05  &            Friel et al. (2010)  &   SF \\
          Be 66  &   3.98  &   12.0  &  -0.48  &  -0.46  &        Villanova et al. (2005)  &   SF \\
          Be 73  &   1.50  &   16.7  &  -0.22  &  -0.24  &          Carraro et al. (2007b)  &    C \\
          Be 75  &   3.00  &   15.6  &  -0.22  &  -0.24  &          Carraro et al. (2007b)  &    C \\
        Blanco1  &   0.1   &    7.9  &   0.4   &  -0.02  &                \cite{Ford2005}  &    W \\
         Cr 110  &   1.70  &    9.7  &   0.03  &   0.03  &          Pancino et al. (2010)  &    B$^c$ \\
        Cr 261   &   6.00  &    6.9  &   0.13  &   0.12  &          Sestito et al. (2008)  &    B \\
         Hyades  &   0.70  &    8.0  &   0.11  &   0.11  &          Carrera et al. (2011)  &   SF \\
         IC 166  &   2.00  &   10.2  &  -0.32  &  -0.32  &           Friel \& Janes (1993)  &   FJ \\
       IC 4651   &   1.68  &    7.1  &   0.11  &   0.15  &         Carretta et al. (2004)  &   SF \\
         King 8  &   0.80  &   11.3  &  -0.40  &  -0.40  &           Friel \& Janes (1993)  &   FJ \\
           M 67  &   4.30  &    8.5  &  -0.01  &   0.01  &         Jacobson et al. (2011b)  &   SF \\
        Mel 66   &   5.33  &    8.9  &  -0.33  &  -0.34  &          Sestito et al. (2008)  &   SF \\
         Mel 71  &   0.2   &   10.3  &  -0.32  &  -0.10  &                \cite{Brown1996}  &    W \\
       NGC~1193  &   4.23  &   11.4  &  -0.22  &  -0.20  &            Friel et al. (2010)  &   SF \\
       NGC~1245  &   1.06  &   10.5  &  -0.04  &  -0.02  &         Jacobson et al. (2011b)  &   SF \\
       NGC~1817  &   1.12  &    9.7  &  -0.16  &  -0.14  &         Jacobson et al. (2011b)  &   SF \\
        NGC~188  &   6.30  &    8.8  &  -0.03  &  -0.01  &         Jacobson et al. (2011b)  &   SF \\
       NGC~1883  &   0.65  &   11.7  &  -0.01  &   0.01  &         Jacobson et al. (2009)  &    C \\
       NGC~1901  &   0.8   &    7.9  &  -0.08  &  -0.08  &                \cite{Carraro2007b}  &    W \\
       NGC~2099  &   0.43  &    9.2  &   0.01  &   0.01  &          Pancino et al. (2010)  &    B \\
       NGC~2112  &   2.50  &    8.6  &  -0.52  &  -0.52  &           Friel \& Janes (1993)  &   FJ \\
       NGC~2141  &   2.45  &   12.0  &   0.00  &   0.02  &         Jacobson et al. (2009)  &   SF \\
       NGC~2158  &   2.00  &   11.5  &  -0.28  &  -0.26  &         Jacobson et al. (2011b)  &    C \\
       NGC~2194  &   0.87  &   10.5  &  -0.08  &  -0.06  &         Jacobson et al. (2011b)  &   SF \\
       NGC~2204  &   2.00  &   11.2  &  -0.23  &  -0.21  &        Jacobson et al. (2011a)  &   SF \\
      NGC~2243   &   4.80  &   10.1  &  -0.42  &  -0.40  &        Jacobson et al. (2011a)  &    B \\
       NGC~2324  &   0.67  &   10.7  &  -0.17  &  -0.18  &        Bragaglia et al. (2008)  &   SF \\
       NGC~2355  &   0.79  &   10.0  &  -0.08  &  -0.06  &         Jacobson et al. (2011b)  &   SF \\
       NGC~2360  &   1.90  &    8.8  &  -0.28  &  -0.28  &           Friel \& Janes (1993)  &   FJ \\
       NGC~2420  &   2.20  &   10.0  &  -0.20  &  -0.18  &         Jacobson et al. (2011b)  &   SF \\
       NGC~2425  &   3.55  &    9.5  &  -0.15  &  -0.13  &         Jacobson et al. (2011b)  &    C \\
       NGC~2477  &   1.00  &    8.3  &   0.07  &   0.06  &        Bragaglia et al. (2008)  &   SF \\
       NGC~2506  &   1.70  &   10.3  &  -0.20  &  -0.16  &         Carretta et al. (2004)  &    B \\
       NGC~2539  &   0.4   &    8.8  &   0.13  &   0.13  &            \cite{Santos2009}  &    W \\
       NGC~2660  &   0.73  &    8.5  &   0.04  &   0.03  &        Bragaglia et al. (2008)  &   SF \\
       NGC~2682  &   5.00  &    8.5  &  -0.09  &  -0.09  &           Friel \& Janes (1993)  &   FJ \\
       NGC~3680  &   4.00  &    7.8  &  -0.16  &  -0.16  &           Friel \& Janes (1993)  &   FJ \\
      NGC~3960   &   0.90  &    7.3  &  -0.12  &  -0.08  &        Bragaglia et al. (2006)  &    B \\
       NGC~5822  &   1.20  &    7.4  &  -0.21  &  -0.21  &           Friel \& Janes (1993)  &   FJ \\
      NGC~6134   &   0.93  &    7.1  &   0.15  &   0.19  &         Carretta et al. (2004)  &    W$^d$ \\
       NGC~6192  &   0.13  &    6.5  &   0.12  &   0.10  &          Magrini et al. (2010)  &    W \\
      NGC~6253   &   3.00  &    6.5  &   0.36  &   0.35  &          Sestito et al. (2008)  &    B \\
       NGC~6404  &   0.50  &    6.1  &   0.11  &   0.09  &          Magrini et al. (2010)  &    C \\
       NGC~6475  &   0.20  &    7.6  &   0.03  &   0.03  &        Villanova et al. (2009)  &    C \\
       NGC~6583  &   1.00  &    5.9  &   0.37  &   0.35  &          Magrini et al. (2010)  &    C \\
       NGC~6791  &   8.00  &    7.6  &   0.42  &   0.42  &          Geisler et al. (2012)  &    C \\
      NGC~6819   &   2.00  &    7.6  &   0.09  &   0.11  &        Bragaglia et al. (2001)  &    B \\
       NGC~6939  &   2.2   &    8.1  &   0.00  &   0.00  &            \cite{Jacobson2007}  &    W \\
       NGC~7142  &   4.04  &    9.2  &   0.13  &   0.15  &         Jacobson et al. (2008)  &   SF \\
        NGC~752  &   1.24  &    8.2  &   0.08  &   0.08  &          Carrera et al. (2011)  &   SF \\
       NGC~7789  &   1.80  &    8.9  &   0.02  &   0.04  &         Jacobson et al. (2011b)  &   SF \\
       Praesepe  &   0.70  &    8.1  &   0.16  &   0.16  &          Carrera et al. (2011)  &   SF \\
           PWM4  &   7.00  &   18.4  &  -0.34  &  -0.34  &             Yong et al. (2012)  &   YON$^e$ \\
     Ruprecht 4  &   0.80  &   12.0  &  -0.09  &  -0.11  &          Carraro et al. (2007b)  &    C \\
     Ruprecht 7  &   0.80  &   13.3  &  -0.26  &  -0.28  &          Carraro et al. (2007b)  &    C \\
       Saurer 1  &   5.00  &   20.1  &  -0.38  &  -0.38  &          Carraro et al. (2004)  &    C \\
           To 2  &   2.50  &   14.9  &  -0.60  &  -0.60  &           Friel \& Janes (1993)  &   FJ \\
\hline\hline
\end{longtable}
\tablefoot{From left to right the columns give the cluster ID, the Galactocentric distance, 
the iron abundance available in the literature, the metallicity re-scaled to current solar 
abundance, the references for the metallicity and the references for ages and distances. The 
acronyms given in column 7 do refer to SF: \cite[][]{Salaris2004,Friel1995}, 
to C (Carraro's group): \citep[][]{Carraro2002,Carraro2005a,Carraro2005b,Carraro2005c,Moitinho2006,Villanova2007,Villanova2009}, 
to B (BOCCE): \citep[][]{Gratton1994,Bragaglia2001,Kalirai2004,Bragaglia2006,Carretta2004,Carretta2005,Carretta2007} 
to W: (WEBDA) and to YON: \cite[][]{Yong2012}.}
\end{center}

\clearpage
\begin{table}
\scriptsize 
\caption{Structural parameters and metallicities of the candidate Cepheid Groups.}\label{tab:islands}
\centering
\begin{tabular}{rccccccc r ccrcc}
\hline
ID     &	x & y & z & $\Rg$ & $\delta\Rg^-$ & $\delta\Rg^+$  & D  & N$_s$ & P & $\sigma$(P) & $\bar{\rho}_{CG}$ & $\Delta$\feh & $\sigma$($\Delta$\feh) \\
       &	(kpc) & (kpc) & (kpc) & (kpc) & (kpc) & (kpc)  & (kpc) &  & (days) & (days) & (kpc$^{-3}$)  & &  \\
\hline
\hline
I   & 0.766 & -1.612 & -0.033 & 6.374 & 0.384 & 0.475 & 1.213 & 20 & 8.69 & 5.17 & 16.8 & -0.039 & 0.010 \\ 
II  &-0.939 & -1.562 & -0.035 & 6.447 & 0.286 & 0.430 & 0.827 & 7  & 8.69 & 3.78 &  4.8 & -0.003 & 0.004 \\ 
III &-1.636 & -0.655 & -0.003 & 7.466 & 0.775 & 0.560 & 1.661 & 37 & 7.48 & 7.35 & 37.7 & -0.063 & 0.010 \\ 
IV  & 0.016 & -0.503 & -0.034 & 7.437 & 0.503 & 0.501 & 1.879 & 52 & 6.91 & 5.20 & 45.7 & -0.009 & 0.008 \\ 
V   & 2.182 & -0.137 &  0.044 & 8.103 & 0.464 & 0.371 & 0.847 & 7  & 8.74 & 5.84 &  8.8 & 0.009 & 0.002 \\ 
VI  & 1.549 &  0.918 & -0.099 & 8.993 & 0.133 & 0.186 & 0.497 & 6  & 5.30 & 1.34 &  4.2 & 0.033 & 0.003 \\ 
VII & 2.608 &  1.167 &  0.050 & 9.473 & 0.396 & 0.190 & 0.654 & 11 & 6.72 & 3.51 &  4.7 & 0.062 & 0.007 \\ 
VIII& 1.732 &  1.515 &  0.014 & 9.613 & 0.219 & 0.261 & 0.641 & 11 & 5.08 & 2.37 &  4.2 & 0.040 & 0.009 \\ 
IX  & 2.511 &  1.777 & -0.008 &10.036 & 0.144 & 0.226 & 0.621 & 7  & 5.38 & 0.79 &  4.4 & 0.046 & 0.006 \\ 
X   & 3.269 &  1.754 & -0.047 &10.231 & 0.187 & 0.230 & 0.449 & 6  & 5.63 & 1.60 &  6.8 & 0.033 & 0.007 \\ 
\hline
\hline
\end{tabular}
\tablefoot{Structural parameters of the ten candidate Cepheid Groups identified in \ref{sec:gradient}. 
From left to right the first column gives the identification number, while the columns 2, 3 and 4 give 
the Galactic coordinates --$x, y, z$-- of the barycenter. Columns 5, 6 and 7 give the Galactocentric 
radius of the barycenter together with the Galactocentric distance of both the inner and the outer 
edge of the Cepheid Group. Column 8 gives the diameter of the CG, while column 9 lists the number of 
Cepheid per CG. Columns 10 and 11 give the mean period and its intrinsic dispersion, while column 12 
gives the density. Column 13 and 14 give the mean residual iron abundance and its intrinsic dispersion.}
\end{table}

\begin{center} 
\scriptsize 
\begin{longtable}{l l rl rl}
\caption{Intrinsic parameters for field and cluster (NGC~1850, NGC~1866) 
Magellanic Cepheids with accurate iron abundances.}\label{tab:amp}\\
\hline
ID &  Period & [Fe/H] & Ref.  & $A_V$ & Ref. \\ 
   &  (days) &        &       & mag   &      \\ 
\hline \hline
  \multicolumn{6}{c}{SMC} \\ 
HV 817 & 18.9234 & -0.83 & \cite{Romaniello2008} & 0.92 & \cite{Caldwell1986,Caldwell2001}\\ 
HV 823 & 31.9154 & -0.81 & \cite{Romaniello2008} & 0.94 & \cite{Soszynski2008,Soszynski2010} \\ 
HV 824 & 65.7658 & -0.74 & \cite{Romaniello2008} & 0.99 & \cite{Soszynski2008,Soszynski2010} \\ 
HV 829 & 85.3100 & -0.75 & \cite{Romaniello2008} & 0.76 & \cite{Soszynski2008,Soszynski2010} \\ 
HV 834 & 73.4514 & -0.64 & \cite{Romaniello2008} & 0.47 & \cite{Soszynski2008,Soszynski2010} \\ 
HV 837 & 42.7563 & -0.82 & \cite{Romaniello2008} & 0.92 & \cite{Caldwell1986,Caldwell2001}\\ 
HV 847 & 27.1019 & -0.76 & \cite{Romaniello2008} & 0.96 & \cite{Soszynski2008,Soszynski2010} \\ 
HV 865 & 33.3426 & -0.85 & \cite{Romaniello2008} & 1.14 & \cite{Caldwell1986,Caldwell2001}\\ 
HV 1365 & 12.4165 & -0.83 & \cite{Romaniello2008} & 0.77 & \cite{Caldwell1986,Caldwell2001}\\ 
HV 1954 & 16.7109 & -0.76 & \cite{Romaniello2008} & 0.79 & \cite{Caldwell1986,Caldwell2001}\\ 
HV 2064 & 33.6512 & -0.64 & \cite{Romaniello2008} & 0.94 & \cite{Soszynski2008,Soszynski2010} \\ 
HV 2195 & 41.7830 & -0.68 & \cite{Romaniello2008} & 0.87 & \cite{Soszynski2008,Soszynski2010} \\ 
HV 2209 & 22.6464 & -0.66 & \cite{Romaniello2008} & 0.61 & \cite{Soszynski2008,Soszynski2010} \\ 
HV 11211 & 21.3796 & -0.82 & \cite{Romaniello2008} & 0.92 & \cite{Caldwell1986,Caldwell2001}\\ 
HV 821  & 127.3140 & -0.79 & \cite{Luck1998} & 1.14 & \cite{Soszynski2008,Soszynski2010} \\ 
HV 824  & 65.8635 & -0.75 & \cite{Luck1998} & 0.99 & \cite{Soszynski2008,Soszynski2010} \\ 
HV 829  & 85.1990 & -0.65 & \cite{Luck1998} & 0.76 & \cite{Soszynski2008,Soszynski2010} \\ 
HV 834  & 73.6390 & -0.66 & \cite{Luck1998} & 0.47 & \cite{Soszynski2008,Soszynski2010} \\ 
HV 11157  & 69.0872 & -0.76 & \cite{Luck1998} & 0.43 & \cite{Soszynski2008,Soszynski2010} \\ 
  \multicolumn{6}{c}{LMC} \\ 
G 458 & 74 & -0.4 & \cite{Luck1992} & 0.48 & van Genderen/\cite{vanGenderen1989,Caldwell2001}\\ 
HV 1003 & 24.4 & -0.75 & \cite{Luck1992} & 1.20 & \cite{Karczmarek2012} \\ 
HV 1013 & 24.0991 & -0.60 & \cite{Romaniello2008} & 0.95 & \cite{Karczmarek2012} \\ 
HV 1023 & 26.6073 & -0.28 & \cite{Romaniello2008} & 0.87 & \cite{Martin1979}\\ 
HV 12452 & 8.7297 & -0.36 & \cite{Romaniello2008} & 0.82 & \cite{Soszynski2008,Soszynski2010} \\ 
HV 12700 & 8.1470 & -0.36 & \cite{Romaniello2008} & 0.56 & \cite{Martin1979}\\ 
HV 2257  & 39.3699 & -0.36 & \cite{Luck1998} & 1.35 & \cite{Karczmarek2012} \\ 
HV 2260 & 12.9420 & -0.37 & \cite{Romaniello2008} & 0.85 & \cite{Soszynski2008,Soszynski2010} \\ 
HV 2294 & 36.5595 & -0.42 & \cite{Romaniello2008} & 1.27 & \cite{Karczmarek2012} \\ 
HV 2337 & 6.8707 & -0.36 & \cite{Romaniello2008} & 0.87 & \cite{Soszynski2008,Soszynski2010} \\ 
HV 2338  & 42.1944 & -0.37 & \cite{Luck1998} & 1.22 & \cite{Karczmarek2012} \\ 
HV 2352 & 13.6144 & -0.48 & \cite{Romaniello2008} & 0.88 & \cite{Karczmarek2012} \\ 
HV 2369 & 48.3059 & -0.63 & \cite{Romaniello2008} & 1.14 & \cite{Karczmarek2012} \\ 
HV 2405 & 6.9183 & -0.28 & \cite{Romaniello2008} & 0.47 & \cite{Soszynski2008,Soszynski2010} \\ 
HV 2447  & 118.3115 & -0.26 & \cite{Luck1998} & 0.67 & \cite{Karczmarek2012} \\ 
HV 2580 & 16.9044 & -0.25 & \cite{Romaniello2008} & 0.98 & \cite{Karczmarek2012} \\ 
HV 2733 & 8.7297 & -0.28 & \cite{Romaniello2008} & 0.43 & \cite{Martin1979}\\ 
HV 2793 & 19.1867 & -0.11 & \cite{Romaniello2008} & 0.84 & \cite{Soszynski2008,Soszynski2010} \\ 
HV 2827 & 78.8860 & -0.36 & \cite{Romaniello2008} & 0.63 & \cite{Karczmarek2012} \\ 
HV 2836 & 17.5388 & -0.18 & \cite{Romaniello2008} & 1.06 & \cite{Martin1979}\\ 
HV 2864 & 10.9901 & -0.20 & \cite{Romaniello2008} & 0.81 & \cite{Martin1979,Caldwell1986}\\ 
HV 2883  & 108.9680 & -0.20 & \cite{Luck1998} & 1.24 & \cite{Martin1979,vanGenderen1983,Freedman1985}\\ 
HV 5497 & 99.3116 & -0.25 & \cite{Romaniello2008} & 0.66 & \cite{Karczmarek2012} \\ 
HV 6093 & 4.7863 & -0.6 & \cite{Romaniello2008} & 0.81 & \cite{Soszynski2008,Soszynski2010} \\ 
HV 877 & 45.0817 & -0.46 & \cite{Romaniello2008} & 0.92 & \cite{Karczmarek2012} \\ 
HV 879 & 36.8129 & -0.14 & \cite{Romaniello2008} & 1.28 & \cite{Karczmarek2012} \\ 
HV 883  & 133.5850 & -0.19 & \cite{Luck1998} & 1.41 & \cite{Karczmarek2012} \\ 
HV 900  & 47.5085 & -0.43 & \cite{Luck1998} & 1.09 & \cite{Karczmarek2012} \\ 
HV 909  & 37.5654 & -0.27 & \cite{Luck1998} & 1.09 & \cite{Karczmarek2012} \\ 
HV 953 & 47.9 & -0.47 & \cite{Luck1992} & 1.08 & \cite{Karczmarek2012} \\ 
HV 971 & 9.2897 & -0.29 & \cite{Romaniello2008} & 1.42 & \cite{Martin1979}\\ 
HV 997 & 13.1522 & -0.22 & \cite{Romaniello2008} & 0.91 & \cite{Soszynski2008,Soszynski2010}\\ 
  \multicolumn{6}{c}{NGC~1850} \\ 
9 & 30.4 & -0.12 & \cite{Sebo1995} & 1.2 & \cite{Sebo1995} \\ 
17 & 18.66 & -0.12 & \cite{Sebo1995} & 0.8 & \cite{Sebo1995} \\ 
110 & 11.858 & -0.12 & \cite{Sebo1995} & 1.0 & \cite{Sebo1995} \\ 
58 & 8.558 & -0.12 & \cite{Sebo1995} & 1.0 & \cite{Sebo1995} \\ 
269 & 7.01 & -0.12 & \cite{Sebo1995} & 0.9 & \cite{Sebo1995} \\ 
341 & 3.593 & -0.12 & \cite{Sebo1995} & 0.8 & \cite{Sebo1995} \\ 
679 & 2.7105 & -0.12 & \cite{Sebo1995} & 0.8 & \cite{Sebo1995} \\ 
  \multicolumn{6}{c}{NGC~1866} \\ 
HV 12197 & 3.14381 & -0.43 & \cite{Mucciarelli2011} & 0.77 & \cite{Welch1991} \\ 
HV 12198 & 3.52279 & -0.43 & \cite{Mucciarelli2011} & 0.65 & \cite{Welch1991} \\ 
HV 12199 & 2.63918 & -0.43 & \cite{Mucciarelli2011} & 0.69  & \cite{Welch1991} \\ 
HV 12200 & 2.7248 & -0.43 & \cite{Mucciarelli2011} & 0.82 & \cite{Welch1991} \\ 
HV 12202 & 3.10112 & -0.43 & \cite{Mucciarelli2011} & 0.55 & \cite{Welch1991} \\ 
HV 12203 & 2.95411 & -0.43 & \cite{Mucciarelli2011} & 0.65 & \cite{Welch1991} \\ 
HV 12204 & 3.43876 & -0.43 & \cite{Mucciarelli2011} & 0.93 & \cite{Welch1991} \\ 
V4 & 3.3157 & -0.43 & \cite{Mucciarelli2011} & 0.25 & \cite{Welch1991} \\ 
V6 & 1.944252 & -0.43 & \cite{Mucciarelli2011} & 0.38 & \cite{Welch1991} \\ 
V7 & 3.453 & -0.43 & \cite{Mucciarelli2011} & 0.46 & \cite{Welch1991} \\ 
V8 & 2.0088 & -0.43 & \cite{Mucciarelli2011} & 0.31 & \cite{Welch1991} \\ 
We 2 & 3.054847 & -0.43 & \cite{Mucciarelli2011} & 0.65 & \cite{Welch1993} \\ 
We 3 & 3.045 & -0.43 & \cite{Mucciarelli2011} & 0.60 & \cite{Welch1993} \\ 
We 4 & 2.8604 & -0.43 & \cite{Mucciarelli2011} & 0.45 & \cite{Welch1993} \\ 
We 6 & 3.289 & -0.43 & \cite{Mucciarelli2011} & 0.50 & \cite{Welch1993} \\ 
We 8 & 3.043 & -0.43 & \cite{Mucciarelli2011} & 0.51 & \cite{Welch1991} \\ 
WS 11 & 3.0544 & -0.43 & \cite{Mucciarelli2011} & 0.40 & \cite{Welch1993} \\ 
WS 5 & 2.8950 & -0.43 & \cite{Mucciarelli2011} & 0.40 & \cite{Welch1993} \\ 
WS 9 & 3.0710 & -0.43 & \cite{Mucciarelli2011} & 0.44 & \cite{Welch1991} \\ 
\hline \hline
\end{longtable}
\tablefoot{From left to right the columns give the period, the iron abundance 
available in the literature and its reference, the visual amplitude and its 
reference.}
\end{center} 

\twocolumn

\end{document}